\documentclass[twocolumn,preprintnumbers,amsmath,amssymb]{revtex4-1}
\usepackage{epsfig}
\usepackage{color}
\usepackage{amsmath}
\usepackage{float}
\usepackage{amsfonts}
\usepackage{amssymb}
\usepackage{multirow}
\usepackage{graphicx,stackengine}
\usepackage{subfig}
\usepackage[percent]{overpic}
\usepackage[english]{babel}
\usepackage{float}
\usepackage{array}
\usepackage{pst-plot}
\usepackage[version=4]{mhchem}
\usepackage{lipsum}
\usepackage{verbatim}
\usepackage{tikz}
\usetikzlibrary{calc,fadings}
\tikzfading[name=fade l,left color=transparent!100,right color=transparent!0]
\tikzfading[name=fade r,right color=transparent!100,left color=transparent!0]
\tikzfading[name=fade d,bottom color=transparent!100,top color=transparent!0]
\tikzfading[name=fade u,top color=transparent!100,bottom color=transparent!0]

\begin{document}
\title{A Deep Potential model for liquid-vapor equilibrium and cavitation rates of water}

\author{Ignacio Sanchez-Burgos$^{1,2}$, Maria Carolina Muniz$^{2}$, Jorge R. Espinosa$^{1,3}$ and Athanassios Z. Panagiotopoulos$^{2,*}$}
\affiliation{
[1] Maxwell Centre, Cavendish Laboratory, Department of Physics, University of Cambridge, J J Thomson Avenue, Cambridge CB3 0HE, United Kingdom. \\
[2] Department of Chemical and Biological Engineering, Princeton University, Princeton, New Jersey 08544, USA. \\
[3] Departamento de Química Física, Facultad de Ciencias Químicas, Universidad Complutense de Madrid, 28040 Madrid, Spain. \\
* = To whom correspondence should be sent.
email: azp@princeton.edu}

\date{\today}

\begin{abstract}

Computational studies of liquid water and its phase transition into vapor have traditionally been performed using classical water models. Here we utilize the Deep Potential methodology ---a machine learning approach--- to study this ubiquitous phase transition, starting from the phase diagram in the liquid-vapor coexistence regime. The machine learning model is trained on \emph{ab initio} energies and forces based on the SCAN density functional which has been previously shown to reproduce  solid phases and other properties of water. Here, we compute the surface tension, saturation pressure and enthalpy of vaporization for a range of temperatures spanning from 300 to 600 K, and evaluate the Deep Potential model performance against experimental results and the semi-empirical TIP4P/2005 classical model. Moreover, by employing the seeding technique, we evaluate the free energy barrier and nucleation rate at negative pressures for the isotherm of 296.4 K. We find that the nucleation rates obtained from the Deep Potential model deviate from those computed for the TIP4P/2005 water model, due to an underestimation in the surface tension from the Deep Potential model. From analysis of the seeding simulations, we also evaluate the Tolman length for the Deep Potential water model, which is (0.091 $\pm$ 0.008) nm at 296.4 K. Lastly, we identify that water molecules display a preferential orientation in the liquid-vapor interface, in which H atoms tend to point towards the vapor phase to maximize the enthalpic gain of interfacial molecules. We find that this behaviour is more pronounced for planar interfaces than for the curved interfaces in bubbles. This work represents the first application of Deep Potential models to the study of liquid-vapor coexistence and water cavitation.

\end{abstract}
\maketitle

%\pagebreak
\section{Introduction}

Water is a fundamental substance, crucial for life and relevant in many environmental, engineering, and
biological processes \cite{chanda2009organic,chaplin2006we,lester2006reaction,ponomarenko2014ultrasonic,wheeler2008transpiration,adhikari2015mechanism,yu2012effect,kumar2010study,ma2011adsorption}. Due to this, the past decades have seen a significant effort devoted to the development of models to reproduce the behaviour of water in computer simulations \cite{abascal2005general,abascal2005potential,mahoney2001quantum,jorgensen1983comparison,horn2005characterization,berendsen1987missing,mahoney2000five,molinero2009water,hasegawa2011polarizable,ahlstrom1989molecular,wang2013systematic,babin2013development,santra2007accuracy}.
%Such models are typically adjusted to reproduce experimentally measured properties, such as coexistence lines between thermodynamic phases or critical points. 
Although some of these models account for flexibility and polarizability \cite{lambros2020good,hasegawa2011polarizable,ahlstrom1989molecular,wang2013systematic,babin2013development}, the most widely employed models for water are rigid and non-polarizable. These include the TIP4P/2005 \cite{abascal2005general}, TIP4P/ICE \cite{abascal2005potential}, SPC/E \cite{berendsen1987missing}, TIP3P \cite{jorgensen1983comparison} and TIP4P-Ew \cite{horn2004development} among others \cite{molinero2009water,mahoney2000five}. These empirical models have parameters obtained by fitting to  experimentally measured properties, such as coexistence lines between thermodynamic phases or critical points. They are frequently used to describe ionic solutions \cite{benavides2016consensus,espinosa2016calculation,jiang2018forward,sanchez2022direct,lamas2022freezing,blazquez2022madrid} and for biomolecular simulations \cite{bergonzo2015improved,smith2015force}, as well as other applications \cite{piaggi2020phase,sanz2013homogeneous,espinosa2016seeding,niu2019temperature}. \\

In contrast to the classic semi-empirical approach to water modelling, \emph{ab initio} models are determined from first principles and therefore do not require fitting to experimental data \cite{friesner2005ab}. Traditionally this approach has not been applied to large systems due to its computational cost \cite{mlynsky2014comparison,gerber2014ab}. Nonetheless, recent advances in Machine Learning (ML) have allowed the development of deep potential generators \cite{zhang2018deep,zhang2020dp} capable of constructing MD potentials based on \emph{ab initio} models. In this work, we use a ML-based model that has successfully recapitulated the different solid phases of water \cite{zhang2021phase}. This model has been constructed based on the SCAN quantum mechanical density functional, which succesfully reproduces several properties of water \cite{sun2015strongly,sun2016accurate}. This approach to \emph{ab initio} based models is efficient enough to carry out simulations with tens of thousands of water molecules in a computationally affordable way \cite{piaggi2022pnas,han2017deep}. \\

Making use of the machinery provided by ML \emph{ab initio} based models, here we study the liquid-vapor coexistence of water by means of Molecular Dynamics (MD) simulations. The liquid-vapor properties of water are well known from experiments \cite{Linstrom}. From the computational side, the TIP4P/2005 non-polarizable rigid water model \cite{abascal2005general} has been highly successful at describing the liquid-vapor coexistence properties, reproducing the experimental phase diagram \cite{vega2006vapor} and surface tension \cite{vega2007surface,mountain2009internally,alejandre2010surface} reasonably well, as well as transport properties \cite{vega2006vapor,abascal2005general,pi2009anomalies}. The TIP4P/2005 model has also been extensively benchmarked in the study of liquid-to-vapor and vapor-to-liquid phase transitions \cite{menzl2016molecular,min2019bubbles,abascal2013homogeneous,joswiak2013size,joswiak2016energetic,gonzalez2015bubble,gonzalez2014detecting}. Therefore, we compare the Deep Potential Molecular Dynamics (DPMD) water model performance with that of TIP4P/2005 in addition to experimental data, when available. \\

To test the DPMD model we first evaluate the liquid-vapor phase diagram and measure properties at equilibrium such as the surface tension or the enthalpy of vaporization. We also focus on the liquid-to-vapor phase transition at negative pressure, a phenomenon known as cavitation. Under negative pressures, water can remain metastable with respect to the vapor (the most stable phase under these conditions) for a finite amount of time before undergoing the phase transition \cite{debenedetti2021metastable}. This results from the fact that the phase transition is an activated process, and requires the formation of a critical bubble, i.e., one that has surmounted a free energy barrier and can continue growing irreversibly without a free energy penalty \cite{kashchiev2000nucleation,sanz2013homogeneous}. This happens because the formation of a bubble intrinsically requires the formation of a liquid-vapor interface, which comes associated with an energetic cost, the surface tension. When the phase transition is initiated by the formation of water bubbles within the liquid bulk and in absence of any surface or external agent, the process is termed homogeneous cavitation.\\

It has been experimentally determined that, at ambient temperature, water can sustain negative pressures of up to $-120$ MPa before transitioning into the vapor phase \cite{green1990water,zheng1991liquids,alvarenga1993elastic,azouzi2013coherent,pallares2014anomalies}. While experiments have determined the cavitation pressure, which is the pressure at which the phase transition is observed, computational studies using the TIP4P/2005 model have been able to compute the nucleation rate, a crucial quantity to characterize the cavitation process. The nucleation rate is defined as the number of critical clusters formed per unit of time and volume. The nucleation rate obtained in previous studies for the TIP4P/2005 model \cite{menzl2016molecular,min2019bubbles,abascal2013homogeneous,gonzalez2015bubble,gonzalez2014detecting} will be used as a reference for our DPMD calculations since there are no reliable experimental data for this quantity. \\

Aside from the nucleation rate, we also compare in this work the nucleation free energy barrier and the Tolman length \cite{tolman1949effect}, a parameter employed to describe the change in surface tension with curvature. Finally, we characterize the orientational distribution of water molecules at the interface. We find that the DPMD model can reproduce well the phase diagram of water, but displays a lower surface tension than experimental results or the TIP4P/2005 model. The nucleation rate is consequently greater for the DPMD model compared to the TIP4P/2005 model. \\

\section{Methods}

\subsection{DPMD Model}

We use the recently developed DPMD model for water \cite{zhang2021phase} to perform simulations in the liquid-vapor coexistence regime. The model was generated using an iterative concurrent learning scheme, deep potential generator \cite{zhang2020dp}, to construct a potential energy landscape based on  SCAN \cite{sun2015strongly}, a non-empirical functional that recapitulates several properties of water \cite{sun2016accurate}, such as molecular geometry and solid structures. The final training set used to construct this model included ice and liquid phases snapshots. In Ref. \cite{zhang2021phase}, the phase diagram for this model was calculated for the different ice phases, reaching a reasonable agreement with experimental data \cite{salzmann2011polymorphism,wagner2011new,brown1966preliminary}. For computational purposes we employ the compressed version of this potential, making use of the scheme developed in Ref. \cite{lu2022dp}. Thanks to this approach, we are capable of reaching a computational performance of 5.2 nanoseconds of simulation time per wall clock day for a system of about 10000 water molecules running with a single 2.8 GHz Intel Ice Lake node using four NVIDIA A100 80GB GPUs and $28$ CPU-cores. Example files of simulations employing this potential are available in the Princeton DataSpace repository  \href{https://doi.org/10.34770/ms7d-wm45}{https://doi.org/10.34770/ms7d-wm45} .

\subsection{Simulation details}

Simulations of the DPMD water model were performed using the LAMMPS package \cite{Plimpton1995FastDynamics}, built with the DeePMD-kit \cite{wang2018deepmd}. Seeding and Direct Coexistence (DC) simulations were performed in the $NVT$ ensemble, keeping the number of particles $N$, system volume $V$, and temperature $T$ constant with the Nose-Hoover thermostat \cite{nosethermo,hooverthermo,hoover1986constant}. Additionally, to compute equations of state and to observe crystallization directly at high supersaturations we performed simulations in the $NPT$ ensemble using the Nose-Hoover barostat \cite{nosethermo,hooverthermo,hoover1986constant}. The equations of motion were integrated using the velocity-Verlet integrator. The simulation timestep was 0.5 fs, and the thermostat relaxation time 0.1 ps. In $NPT$ simulations, the barostat relaxation time was 1 ps. \\

For the DC simulations, a system size of 1024 molecules was used and the density profiles were obtained with at least 10 ns of sampling. Coexistence densities were obtained by fitting the density profile to the following expression:

\begin{equation}
    \rho(z) = \frac{\rho_l+\rho_v}{2} - \frac{\rho_l-\rho_v}{2} \mathrm{tanh} \bigg(\frac{z-z_0}{d}\bigg) \label{eq:1}
\end{equation}
where  $\rho_l$ and $\rho_v$ are the coexistence liquid and vapor densities, respectively, $z_0$ the position of the interface, and $d$ its thickness. \\

The surface tension was calculated from DC simulations at each temperature according to the Kirkwood-Buff equation \cite{kirkwood1949statistical}:

    \begin{equation}
        \gamma = \frac{L_z }{2}[\langle P_{zz} \rangle -0.5(\langle P_{xx} \rangle+ \langle P_{yy} \rangle)]
        \label{kirkw}
    \end{equation}
where $P_{ii}$ are the diagonal components of the pressure tensor and $L_z$, the box length in the elongated dimension, perpendicular to the slab interfaces. \\

We also performed simulations with the TIP4P/2005 water model \cite{abascal2005general} using the GROMACS 4.6.7 Molecular Dynamics package \cite{bekker1993gromacs} in the $NPT$ and $NVT$ ensembles, keeping temperature constant with the velocity-rescale thermostat \cite{bussi2007canonical} and pressure constant (for $NPT$ simulations) with the Parrinello-Rahman barostat \cite{parrinello1981polymorphic}. In GROMACS we integrated the equations of motion using the Leap-Frog integrator \cite{hockney1974quiet}. The simulation timestep was 2 fs, and the thermostat and barostat relaxation times were 0.75 and 2 ps, respectively. We set the cut-off of both dispersion interactions and the real part of the electrostatic interactions at 12 \r{A}. Long-range Coulombic interactions were treated with the Particle-Mesh Ewald (PME) solver \cite{darden1993particle,essmann1995smooth}. We kept the O-H bond length (0.9572 \r{A}) and H-O-H angle (104.52$^o$) values constant with the LINCS algorithm \cite{hess1997lincs}. With this model we reached a computational performance of 40 nanoseconds of simulation time per wall clock day for a system of about 10000 water molecules running with Intel(R) Xeon(R) Platinum 8368Q CPU @ 2.60GHz, using 32 CPUs in paralel.

\subsection{Seeding and Classical Nucleation Theory}
\label{cnt}

Seeding is a method that consists of using Classical Nucleation Theory (CNT) in combination with MD simulations \cite{sanz2013homogeneous,espinosa2016seeding}. More specifically, we use the NVT seeding approach \cite{rosales2020seeding}, in which a cluster (in this case a bubble) close in size to the critical one is artificially inserted into the system, then spontaneously equilibrated into the critical size and tracked along time. With this approach, a critical bubble can be characterized for long timescales because the maximum in free energy barrier in a nucleation process represents a minimum in the Helmholtz free energy landscape in the canonical ensemble \cite{montero2020interfacial,min2019bubbles}. Therefore, more precise measurements can be made compared to seeding in the NPT ensemble, where the bubble will rapidly either shrink or grow \cite{rosales2020seeding}. This method is suitable to measure nucleation rates along isotherms, since the pressure at which the cluster is critical cannot be known \emph{a priori}, and is obtained from the simulations. \\

CNT \cite{volmer1926keimbildung,becker1935kinetische} is a theoretical framework that describes nucleation processes under saturation conditions. It can be used to obtain the free energy barrier, interfacial free energy and nucleation rate. The limitations of quantitatively characterizing nucleation rates using CNT are due to assumptions inherent in the theory \cite{sear2012non,merikanto2007origin,cacciuto2004breakdown,horsch2008modification,moroni2005interplay,leoni2021nonclassical}. Despite these potential limitations, multiple studies position CNT as a powerful tool to estimate free energy barriers and nucleation rates for phase transitions \cite{sanz2013homogeneous,espinosa2016seeding,espinosa2018homogeneous,knott2012homogeneous,lamas2021homogeneous,saika2011nucleation,richard2018crystallization,separdar2021molecular,sanchez2021parasitic,sanchez2021fcc,
filion2010crystal,fiorucci2020effect,piaggipnas}, including cavitation \cite{baidakov2022stability,sanchez2020equivalence,rosales2020seeding,menzl2016molecular}. According to CNT \cite{volmer1926keimbildung,blander1975bubble}, the nucleation rate ($J$) can be computed as

\begin{equation}
    J=\rho_l\sqrt{\frac{2\gamma}{\pi m}}exp\left( \frac{-\Delta G^*}{k_BT} \right)
    \label{eq1}
\end{equation}
where $\rho_l$ represents the density of the liquid phase, $\gamma$ the liquid-vapor surface tension, $m$ the mass of water, $\Delta G^*$ the free energy barrier for nucleation, $k_B$ the Boltzmann's constant and $T$ the temperature. Within the CNT framework, the free energy barrier is obtained as

\begin{equation}
    \Delta G^*=\frac{4}{3}\gamma \pi R_c
    \label{eq2}
\end{equation}
where $R_c$ is the critical bubble radius. Additionally, we obtain the interfacial free energy from Laplace's equation as

\begin{equation}
    \gamma =\frac{R_c \Delta P}{2}
    \label{eq3}
\end{equation}
where $\Delta P$ is the pressure difference between the vapor and liquid phases. This approach provides more reliable estimations than assuming the capillarity approximation (i.e. inserting the surface tension at planar interface and coexistence conditions into the CNT) \cite{bal2022extending,diemand2014direct,rosales2020seeding,sanchez2020equivalence}. Combining Eqs. \ref{eq1}, \ref{eq2} and \ref{eq3}, we reach the final equation for $J$:

\begin{equation}
    J=\rho_l \sqrt{\frac{R_c \Delta P}{\pi m}} exp\left(\frac{-4 \pi R_c^2 \Delta P}{3k_BT}\right)
    \label{eq4}
\end{equation}
To summarize, in order to compute $J$ we require the pressure difference between the liquid and the vapor phases, and the critical radius of the bubble at the corresponding thermodynamic conditions of $P$ and $T$. Although the difference in pressure can be computed in principle \cite{rosales2020seeding,bal2022extending}, in this study the pressure inside the bubbles is $\sim$0 \cite{vega2006vapor}, therefore $\Delta P$ can be easily estimated as $-P_{liq}$, which is directly obtained through the virial expression. We additionally confirmed that the virial pressure obtained in the system containing a bubble matches that of the bulk liquid (Fig. S1), as recently shown \cite{montero2020young}. \\

\begin{figure}
    \begin{flushleft} (a)  \end{flushleft} \includegraphics[width=\linewidth]{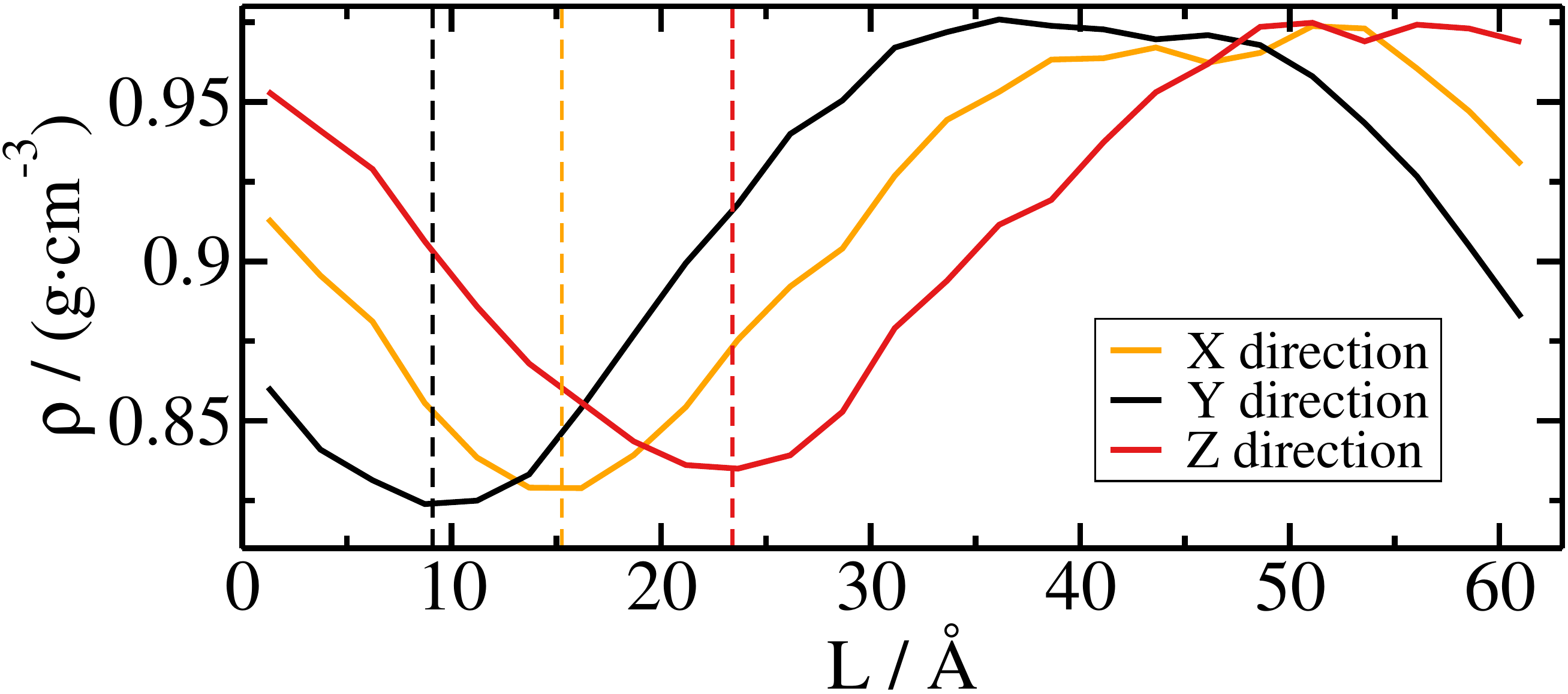}
    \begin{flushleft} (b)  \end{flushleft} \includegraphics[width=\linewidth]{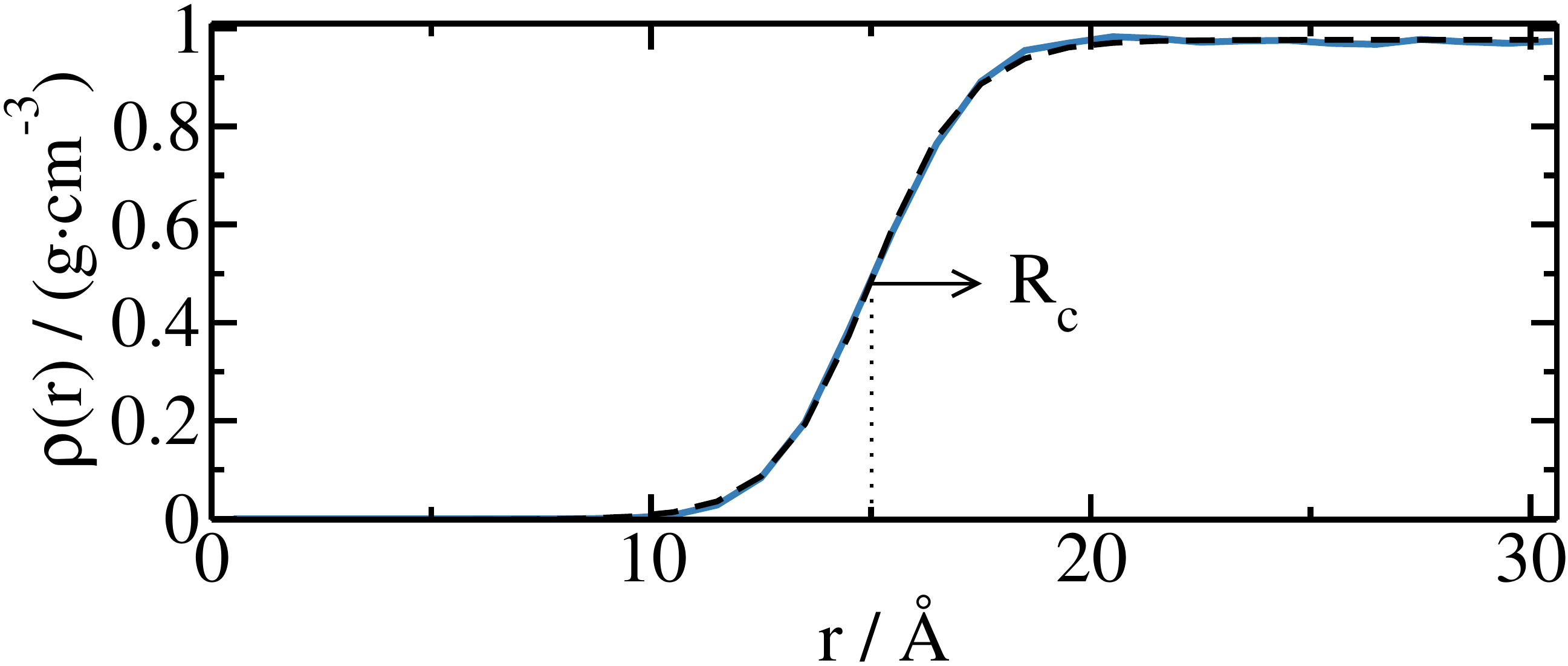}
    
    \caption{(a) Density profiles along the three cartesian directions. Vertical dashed lines depict the location of the minimum density, which corresponds to
    the center of the bubble in each direction. In these density profiles we averaged the density of each point with 2 other neighbouring points in order to make the curves smoother. (b) Radial density profile calculated from the center of the bubble. The blue curve indicates the density calculated at each distance while the black dashed curve is the fit of the blue curve to Eq. \ref{tanheq}, from which we obtained the critical radius R$_c$.}
    \label{figparameter}
\end{figure}

Lastly, the critical radius, $R_c$ was obtained employing a local order parameter. Although multiple parameters have been proposed to track the size of a simulated bubble \cite{meadley2012thermodynamics,kusaka1999identifying,abascal2013homogeneous,gonzalez2015bubble,gonzalez2014detecting,menzl2016molecular}, here we adopted the 'equidensity' criterion \cite{angelil2014bubble}, which was shown to provide the surface of tension radius (i.e. the radius that, when inserted into Laplace's equation provides a consistent value of $\gamma$) for the Lennard-Jones system \cite{rosales2020seeding,sanchez2020equivalence}. As illustrated in Figure \ref{figparameter}(a), the center of the bubble is first calculated through the minimum in the density profile along the 3 cartesian directions. Afterwards, a radial density profile from the bubble center is computed, in which the critical radius ($R_c$) corresponds to the point in which the density equals the average of the liquid and vapor densities (Fig. \ref{figparameter}(b)). This point is found via non-linear fitting to the equation

\begin{equation}
    \rho (r) = \frac{\rho _l + \rho _v }{2} + \frac{\rho _l - \rho _v}{2} tanh \left( \frac{r-R_c}{\alpha} \right)
    \label{tanheq}
\end{equation}
where $r$ is the distance from the bubble center, and $\alpha$ is a fitting parameter. We confirmed that, as assumed by CNT, the bubbles have a close-to-spherical shape (Figure S2).

\section{Results and discussion}

%shift
It is important to note that all DPMD data shown in this work are shifted by $40$ K, so that the simulations for a given temperature have been performed at $40$ K higher than the reported one in the presented figures. Similar shifts in temperature have been performed for AIMD simulations using SCAN \cite{mohanchen2017aimdwater} and in other works using SCAN-based ML models \cite{yaoyi2020scanshift, tom2020signatureLLT, piaggi2021phaseequilibrium}. The rationale for the shift was originally attributed to nuclear quantum effects, but it is likely mainly due to the limitations of the density functional itself. SCAN is known to overestimate the strength of the hydrogen bond \cite{yaoyi2020scanshift}. In this work, the shift in temperature was adjusted by calculating the mean square error between the coexistence vapor densities predicted by the model and the experimental ones, considering different values for the shift. The value for the shift was iteratively modified until we obtained the one that gave the minimum mean square error, which was $40$ K. In all plots and tables that follow the results of the DPMD model have this shift already applied.  \\

\subsection{Vapor-liquid equilibrium in the DPMD model}

\begin{figure*}[ht!]
    \centering
    \begin{tabular}{ccc}
        (a) & (b) & (c) \\
        \begin{overpic}[width=0.305\linewidth]{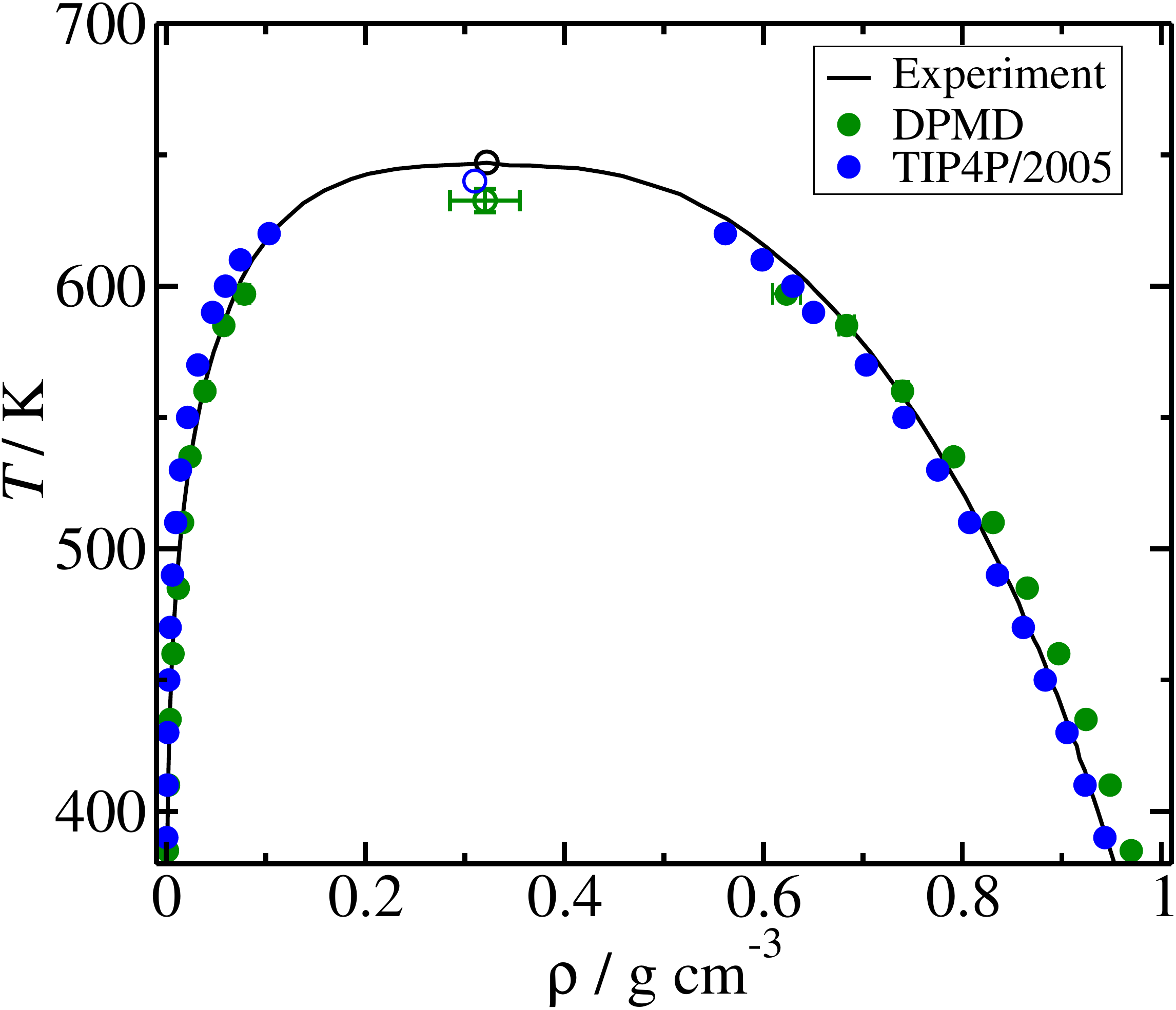}
     \put(17,14.5){\includegraphics[scale=0.08,angle=90]{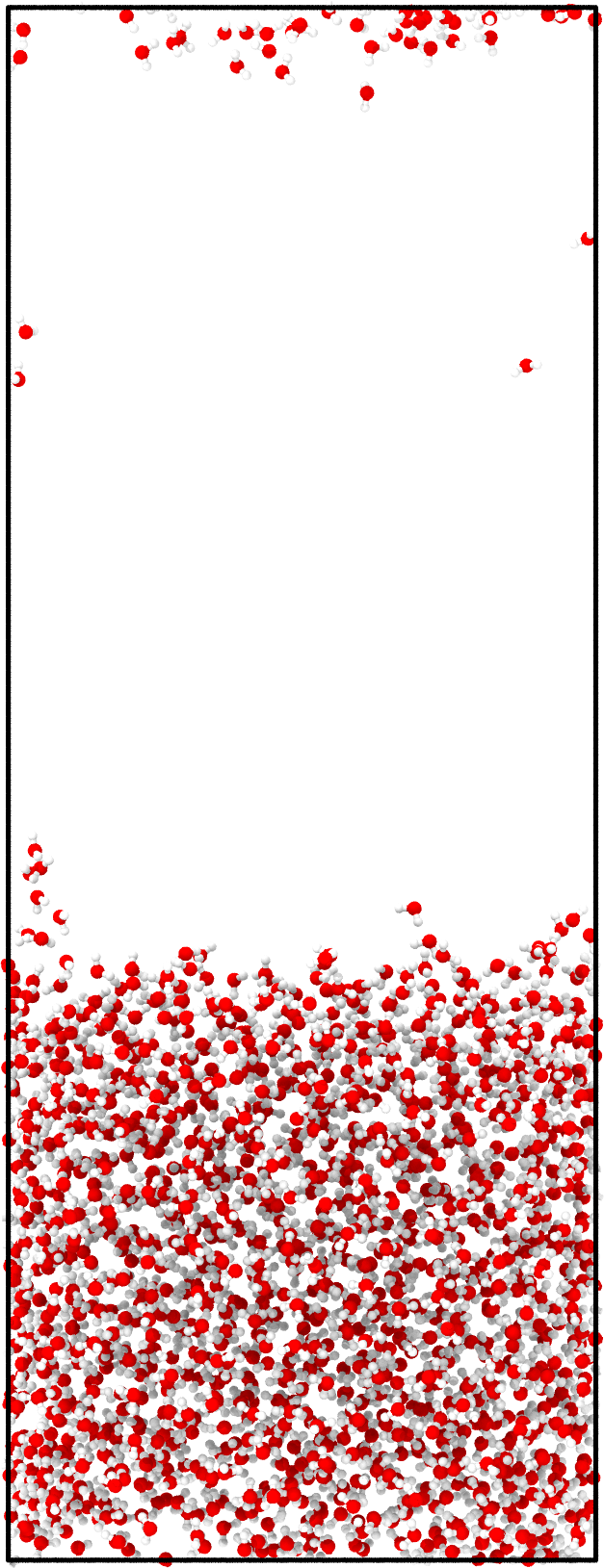}}  
  \end{overpic} &
        
        \includegraphics[width=0.32\linewidth]{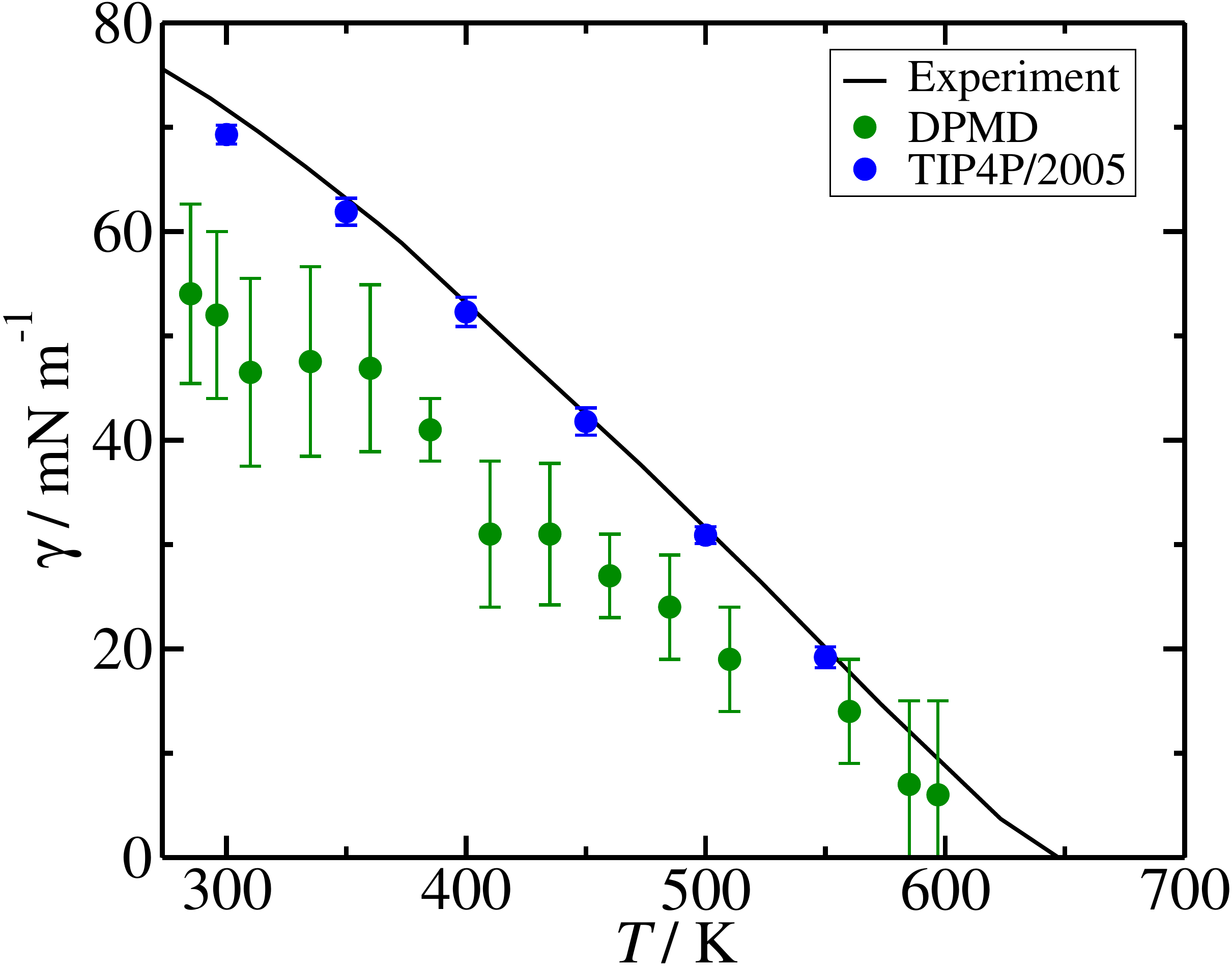} & \includegraphics[width=0.337\linewidth]{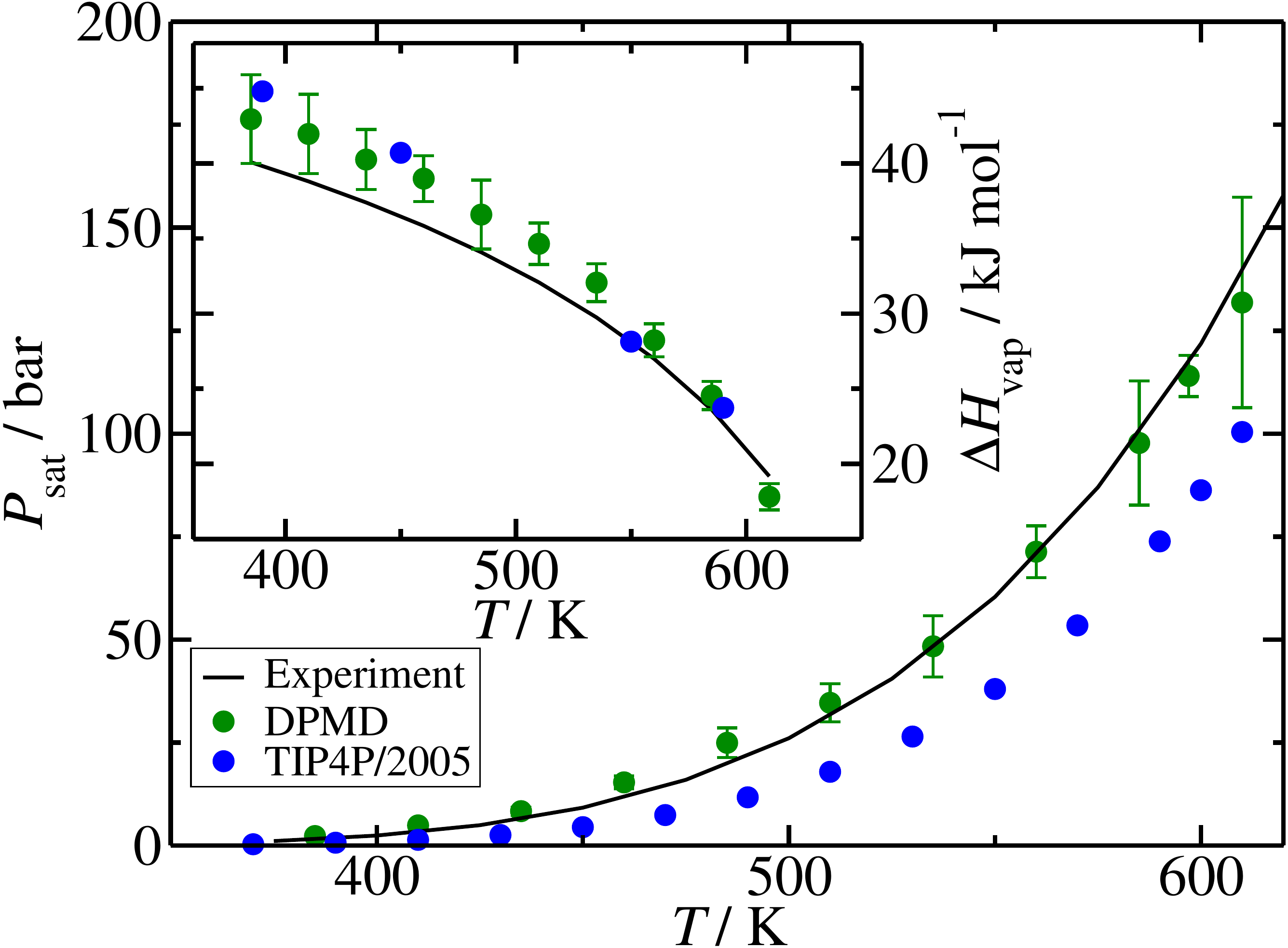}
    \end{tabular}
    \caption{(a) Phase diagram in the $T$–$\rho$ plane of the DPMD model (green), TIP4P/2005 model (blue) \cite{vega2006vapor} and experimental measurements of water (black) \cite{Linstrom}. Filled circles represent the vapor and liquid densities, estimated from the averaged bulk densities from DC simulations. Inset: Snapshot of a DC simulation performed at 385 K with the DPMD model, rendered making use of Ovito software \cite{stukowski2009visualization}. (b) Liquid-vapor interfacial free energy ($\gamma$) as a function of temperature for the DPMD model (green), and comparison with the TIP4P/2005 model (blue) \cite{vega2007surface} and experimental values (black) \cite{Linstrom}. (c) Vapor saturation pressure ($P_{sat}$) as a function of temperature for the DPMD model (green), TIP4P/2005 (blue) \cite{vega2006vapor} and comparison with experimental values (black) \cite{Linstrom}. Inset: Enthalpy of vaporization ($\Delta H_{vap}$) as a function of temperature for DPMD, TIP4P/2005 \cite{vega2006vapor} and experimental values \cite{Linstrom} as indicated in the legend.}
    \label{figure1}
\end{figure*}

%introduce DC and the phase diagram

To test the DPMD model in the liquid-vapor regime, we begin by computing the phase diagram. We do so using simulations in the canonical ensemble, in which both the liquid and vapor phases coexist. In Figure \ref{figure1}(a) (inset) we show a snapshot of a typical DC simulation box. In Figure \ref{figure1}(a) we plot the temperature against density phase diagram of the DPMD model (green points), where the filled points represent the densities directly obtained from DC simulations (2 at each temperature). We include results for the TIP4P/2005 model \cite{abascal2005general,vega2006vapor} (blue circles), as well as experimental data \cite{Linstrom} (black line). We can observe that for the DPMD model the density of the liquid branch is slightly higher than the experimental results at low temperatures ($<$500K), but matches well at higher ones. The critical point is estimated through the universal scaling law of coexistence densities near a critical point \cite{rowlinson2013molecular}, and the law of rectilinear diameters \cite{zollweg1972law}:

\begin{equation}
    \big(\rho_l(T)-\rho_v(T)\big)^{3.06}=d\bigg(1-\frac{T}{T_c}\bigg)
    \label{scaling}
\end{equation}
and 

\begin{equation}
(\rho_l(T)+\rho_v(T))/2=\rho_c+s_2(T_c-T)
    \label{rectilinear}
\end{equation}
where $\rho_l$ and $\rho_v$ refer to the coexisting densities of the liquid and vapor phases respectively, $\rho_c$ is the critical density, $T_c$ is the critical temperature, and $d$ and $s_2$ are fitting parameters. The critical temperature obtained for the DPMD model is of 632.6.6 K, which is lower than the experimental one by 14.5 K. \\

%Surface tension
Moreover, we compute the liquid-vapor surface tension for the DPMD model at different temperatures from the DC simulations. This quantity can be directly estimated according to Eq. \ref{kirkw}. As can be seen in Figure \ref{figure1}(b), the DPMD model provides lower values of $\gamma$ than both the TIP4P/2005 model \cite{vega2007surface} and experimental measurements \cite{Linstrom}. From DC simulations we also calculate the saturation pressure ($P_{sat}$) as a function of temperature. $P_{sat}$ is obtained as the component of the pressure tensor normal to the interface. We plot it in Figure \ref{figure1}(c), compared to the values obtained from TIP4P/2005 (blue) \cite{vega2006vapor} and experiments (black) \cite{Linstrom}. This quantity closely matches with experimental measurements, while the TIP4P/2005 model underestimates it, which is a natural consequence of the way the DPMD model was shifted to match the vapor densities. We note that the temperature shift was applied in order to obtain a better match of the vapor phase behaviour, nonetheless this shift affects negatively on the surface tension prediction. With no temperature shift, the surface tension of the DPMD model matches the experimental one at temperatures above 450 K, and only underestimates it by $\sim 5 \%$ at T $<$ 450 K.\\

We estimated the enthalpy of vaporization ($\Delta H_{vap}$) from independent bulk simulations of both liquid and vapor phases. For this, we perform canonical simulations at the equilibrium density of the given phase, which was previously obtained from DC simulations. From each simulation the enthalpy is directly obtained as $H=U-PV$, where $U$ is the internal energy. The enthalpy of vaporization is simply calculated as $\Delta H_{vap}=H_{vapor}-H_{liquid}$ for every temperature. The values of $\Delta H_{vap}$ are plotted against temperature in Figure \ref{figure1}(c) (inset), along with values from TIP4P/2005 \cite{vega2006vapor} and experiments \cite{Linstrom}. This quantity slightly deviates from the experimental values at low temperatures ($<550$ K), but matches at higher temperatures. Similarly, the TIP4P/2005 model matches the experimental trend at $T>500K$, and overestimates it at lower $T$. \\

In summary, the DPMD model describes the liquid-vapor coexistence properties after applying the temperature shift of 40 K reasonably well. Some discrepancies may arise from the fact that this model has been trained on solid and liquid data only \cite{zhang2021phase}, but the main source of differences from experimental data are limitations in accuracy for the SCAN density functions used to train the model. . The biggest difference with experimental values is found in the surface tension, which is underestimated by $\sim$20 \%. Other than this discrepancy, the phase diagram, saturation pressure, and enthalpy of vaporization are well described using the DPMD model. In addition to the plots in Figure \ref{figure1}, we provide the equilibrium data of the DPMD model in Table \ref{table1}.

\begin{table}[h]
    \centering
    \begin{tabular}{c|c|c|c|c}
        $T$ (K) & $\rho _l$ (g·cm$^{-3}$) & $\rho _v$ (g·cm$^{-3}$) & $P_{sat}$ (bar) & $\gamma$ (mN·m$^{-1}$) \\ \hline
        385 & 0.9697(5) & 0.0011(3) & 2.3(3) & 41(3) \\
        410 & 0.9484(8) & 0.0022(4) & 4.9(5) & 31(7) \\
        435 & 0.9242(6) & 0.0037(6) & 8.3(9) &  31(7)\\
        460 & 0.8969(8) & 0.0067(9) & 15(1) & 27(4)\\
        485 & 0.865(3) & 0.012(4) & 25(3) & 24(5)\\
        510 & 0.831(2) & 0.016(2) & 35(4) & 19(5)\\
        535 & 0.791(1) & 0.024(1) & 48(7) & 15(7)\\
        560 & 0.740(6) & 0.039(5) & 71(6) & 14(5)\\
        585 & 0.684(8) & 0.058(4) & 98(15) & 7(8)\\
        597 & 0.62(1) & 0.079(5) & 114(5) & 6(9)
    \end{tabular}
    \caption{Data for the liquid ($\rho _l$) and vapor ($\rho _v$) densities, saturation pressure ($P_{sat}$) and surface tension ($\gamma$) as a function of temperature for the DPMD model. The numbers in parenthesis depict the uncertainty of our measurements, and apply to the numeral left of themselves, for instance 41(3) stands for 41$\pm$3.}
    \label{table1}
\end{table}

\subsection{Bubble nucleation}
\label{cntsec}

After establishing the equilibrium properties of the DPMD water model, we proceeded to investigate its cavitation. Although some experimental studies of water cavitation have been conducted \cite{green1990water,zheng1991liquids,alvarenga1993elastic,azouzi2013coherent,pallares2014anomalies}, it is difficult to establish a direct comparison due to the lack of measurements of the nucleation rate. Menzl \emph{et al.} \cite{menzl2016molecular} performed a nucleation study utilizing Umbrella Sampling (US) calculations \cite{torrie1974monte} for the TIP4P/2005 model, in which the nucleation free energy barrier and the nucleation rate were reported, without comparisons to experimental data. Here, we employ the NVT seeding technique at the same temperature (296.4 K) as in Ref. \cite{menzl2016molecular} for both the TIP4P/2005 (to establish the validity of our methods) and DPMD models (to provide new data for this ab-initio based model). \\

\begin{figure*}[ht!]
    \centering
    \begin{tabular}{ccc}
        (a) & \ & (b) \\
        \includegraphics[width=0.48\linewidth]{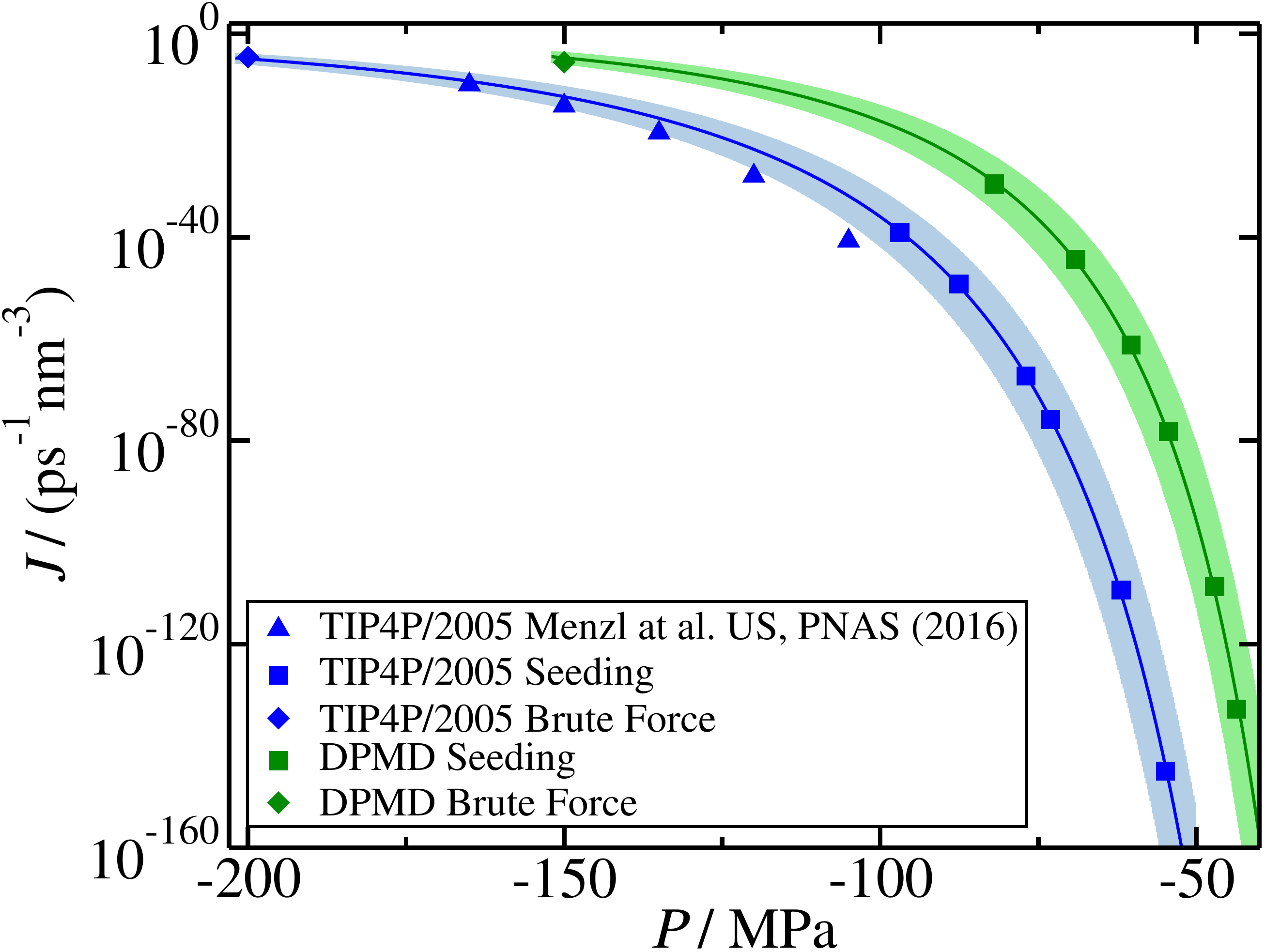} & & \includegraphics[width=0.48\linewidth]{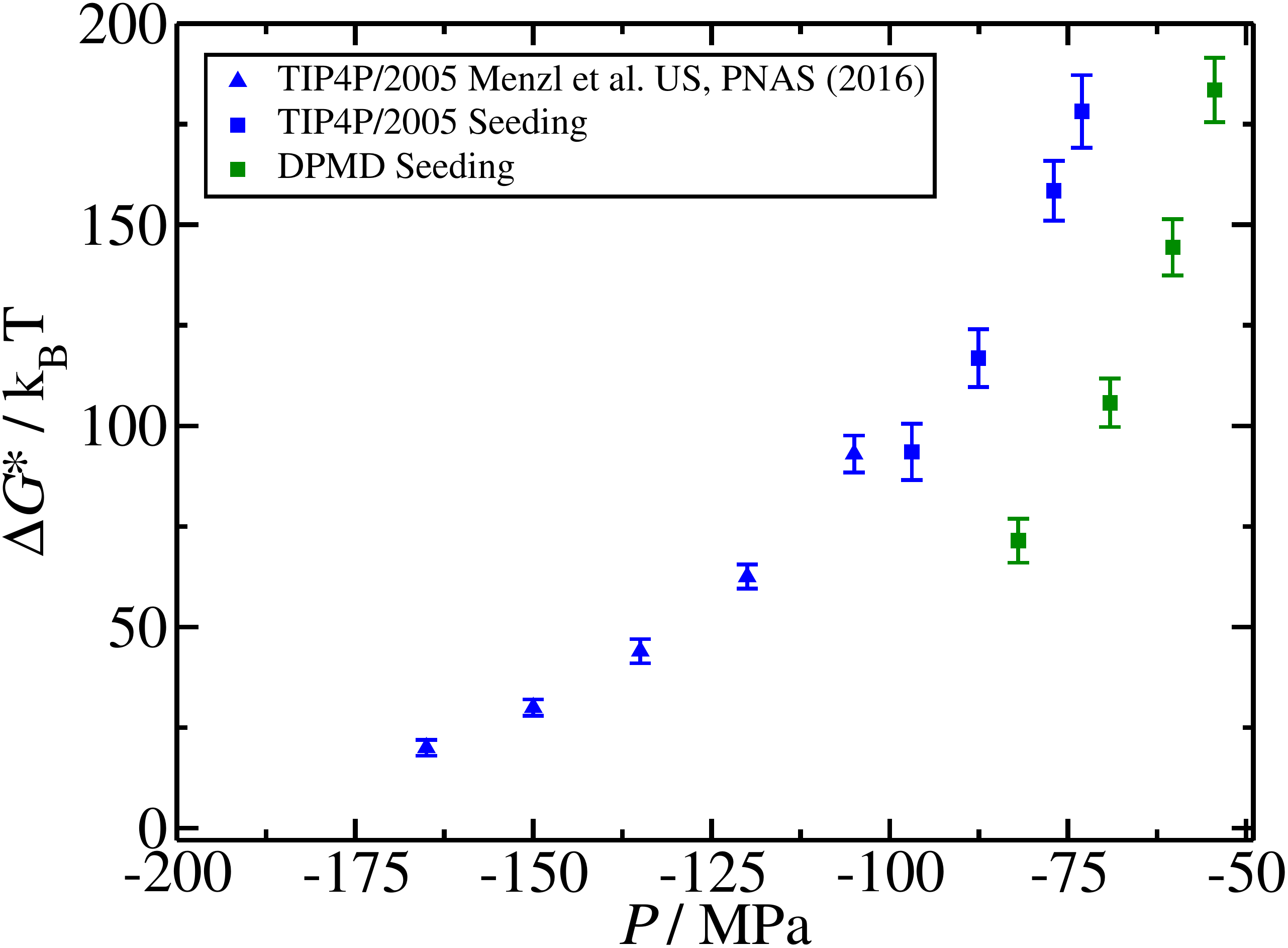}
    \end{tabular}
    \caption{(a) Nucleation rate ($J$) as a function of pressure ($P$) for TIP4P/2005 (blue) and DPMD (green) cavitation at 296.4 K, including data from Ref. \cite{menzl2016molecular}. Continuous lines are obtained by linearly fitting the surface tension ($\gamma$) as a function of pressure, and then inserting such $\gamma$ into Eqs. \ref{eq3} and \ref{eq4}. The shaded region is obtained in the same way but making use of the upper and lower bounds of the surface tension error and, therefore, represent the error limits in $J$ (b) Free energy barrier for bubble nucleation as a function of pressure at 296.4 K for TIP4P/2005 (blue) and DPMD (green) models, estimated from CNT (Eq. \ref{eq2}).}
    \label{figure2}
\end{figure*}

We prepared various systems in which we artificially generated a cavity of a given size, starting from a bulk liquid configuration. As detailed in Section \ref{cnt}, the system spontaneously evolves and equilibrates into a state in which there is a critical bubble that remains stable over time, due to the fact that in the canonical ensemble, a critical bubble represents a local minimum in the Helmholtz free energy landscape \cite{montero2020interfacial,min2019bubbles}. Once equilibrated, we measured the system pressure by means of the virial expression \cite{thompson2009general} which corresponds to the liquid phase pressure \cite{rosales2020seeding}. To track the critical radius, we made use of our order parameter (see Section \ref{cnt}). We repeated this process for each configuration, and then averaged over more than 500 independent radial density profiles for the calculation of the radius. \\

We used our data for $\Delta P$ and $R_c$ along with Eq. \ref{eq4} to compute $J$, which is plotted in Figure \ref{figure2}(a) (blue and green squares for TIP4P/2005 and DPMD respectively) against $P$. It can be seen that there is agreement within the simulation uncertainties between the US and seeding simulations for the TIP4P/2005 model. We also include a continuous line for each model, which represents a fit to the CNT equation, in which we linearly fit $\gamma$ against $P$, and insert values from such fit to solve Eq. \ref{eq4}. The uncertainty is estimated from the standard deviation of the radius between different independent configurations. \\
%The uncertainty in the determination of the pressure is negligible in comparison to that in the determination of the critical radius. The error is therefore derived from the uncertainty in the determination of the critical radius. \\

We computed $J$ at higher superstreching conditions (green and blue diamonds in Fig. \ref{figure2}) through "brute force" simulations. In these simulations, we observed the metastable bulk liquid under high superstreching conditions in the NPT ensemble for a sufficient time before spontaneous cavitation takes place. Then, $J$ is calculated as

\begin{equation}
    J=\frac{1}{<t>V}
    \label{bf}
\end{equation}
where $<t>$ is the average time required for cavitation to occur and $V$ is the volume of the metastable liquid phase. The onset of cavitation can easily be identified with a sudden and sharp change in properties such as the simulation box volume or the potential energy. From these simulations we obtain $J=2.52$·$10^{-6}$ ps$^{-1}$nm$^{-3}$ at $P=-150$ MPa for the DPMD model, and $J=2.44$·$10^{-5}$ ps$^{-1}$nm$^{-3}$ and $P=-200$ MPa for the TIP4P/2005 potential. These results are also shown in Figure \ref{figure2}(a), and match with the trend of US and seeding simulations. This result, in addition to the agreement with the US calculations from Menzl \emph{et al.}, provides confidence in the validity of the results obtained using CNT. \\

In addition to the nucleation rate, we also obtained the free energy barrier ($\Delta G^*$) which can be estimated from Eq. \ref{eq2}. This quantity is the main output from US simulations \cite{menzl2016molecular}. In Figure \ref{figure2}(b) we compare the calculated free energy barriers for the DPMD (green) and TIP4P/2005 (blue) models, also including data from Ref. \cite{menzl2016molecular}. As expected, we find good agreement between seeding and US calculations as for the nucleation rates. \\ 

\begin{figure}[h]
    \centering
    \includegraphics[width=\linewidth]{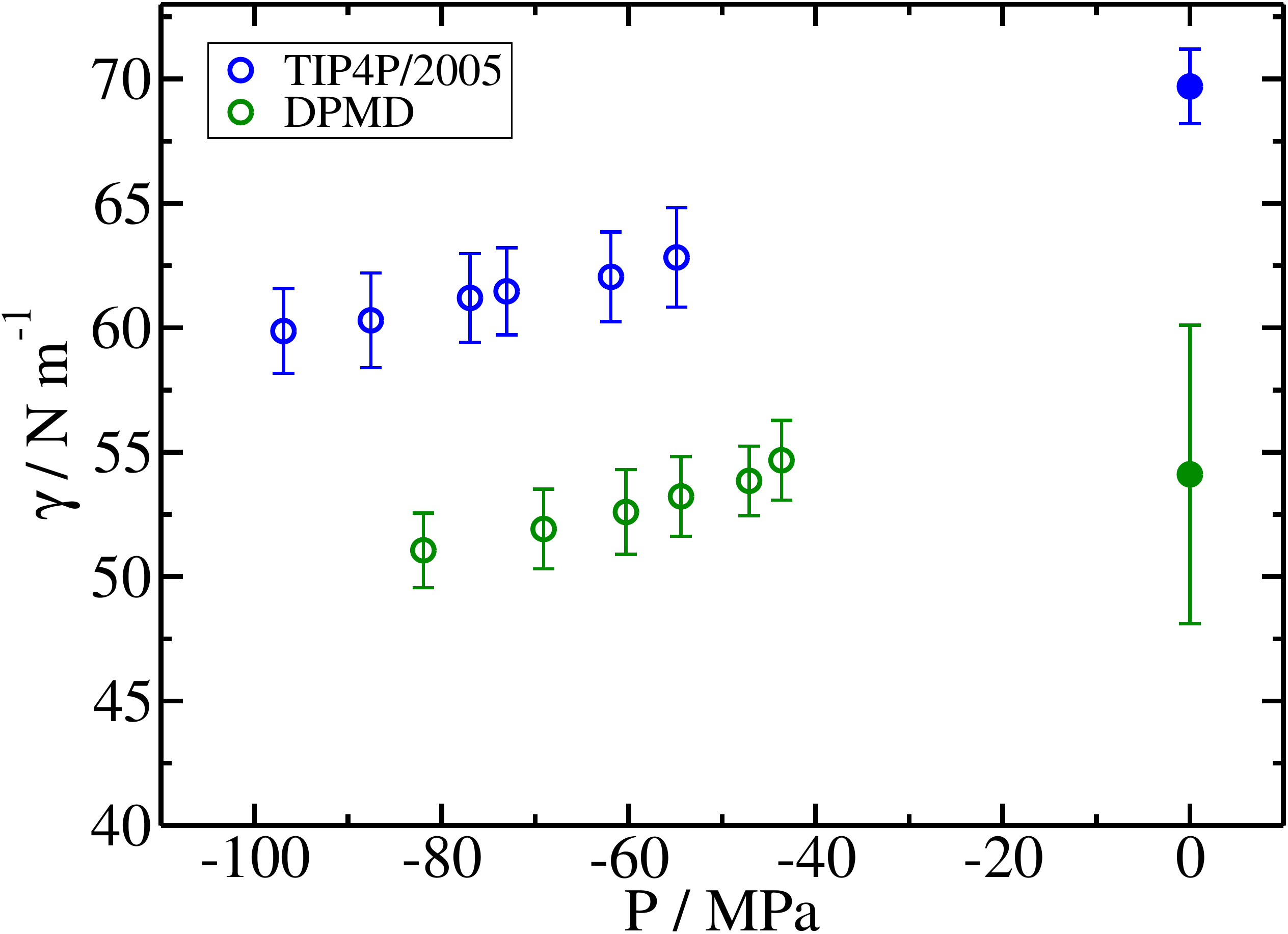}
    \caption{Liquid-vapor surface tension ($\gamma$) as a function of pressure at 296.4 K for TIP4P/2005 (blue) and DPMD (green). Empty circles are obtained from cavitation simulations (and therefore with a curved interface) by means of Laplace's equation (see Section \ref{cnt}), while filled points were estimated from DC simulations as in Fig.\ref{figure1}. }
    \label{figure3}
\end{figure}

It can be seen in Figure \ref{figure2}(a) that the DPMD model returns nucleation rates many orders of magnitude greater than the TIP4P/2005 potential, outside the uncertainty bounds, despite the close resemblance of the phase diagrams from the two models (Fig. \ref{figure1}(a)). This difference is also present for the free energy barrier (Fig. \ref{figure2}(b)), with the TIP4P/2005 model possessing a higher free energy barrier. This is likely the crucial factor behind its lower nucleation rate. In order to understand the differences between the two models we also compared the change of the surface tension with curvature for both models: In Figure \ref{figure3} we plot $\gamma$ as a function of $P$, where filled points correspond to the value obtained at the coexistence pressure from DC simulations, while the empty points depict the value of $\gamma$ obtained through Laplace's equation (Eq. \ref{eq3}) from our seeding simulations. From this analysis we observe that the surface tension is significantly lower for the DPMD model, not only under coexistence conditions, but also in the cavitation regime. This directly points towards the surface tension being the decisive factor behind the quantitative difference in $J$ and $\Delta G^*$ between the DPMD and TIP4P/2005 models. In Table \ref{tab_seeding} we detail the different quantities playing a role in Eqs. 2-4. Since the kinetic prefactor in the calculation of $J$ is of the same order of magnitude in all cases, we can conclude that the different nucleation rates between the TIP4P/2005 and DPMD models arises from a quantitative difference in the surface tension, which is lower for the DPMD.

\begin{table*}[]
    \centering
    \begin{tabular}{c|c|c|c|c|c|c}
        $P$ (MPa) & $R_c$ (nm) & $\rho _l$ (g·cm$^{-3}$) & N$_T$ & $\gamma$ (N·m$^{-1}$) & $\Delta G^*$/k${_B}$T & log$_{10}$($J$ / (ps$^{-1}$·nm$^{-3}$)) \\ \hline
        \multicolumn{7}{c}{DPMD} \\ \hline
        -81.9 & 1.25 & 0.965 & 7527 & 51.0 & 71.5 & -29.5 \\
        -69.1 & 1.50 & 0.972 & 7386 & 51.9 & 105.7 & -44.4 \\
        -60.3 & 1.74 & 0.977 & 7161 & 52.6 & 144.4 & -61.2 \\
        -54.4 & 1.96 & 0.980 & 6888 & 53.2 & 183.5 & -78.2 \\ 
        -47.1 & 2.29 & 0.984 & 10324 & 53.8 & 253.7 & -108.6 \\ 
        -43.7 & 2.50 & 0.986 & 9825 & 54.7 & 309.2 & -132.7 \\ 
        \hline
        \multicolumn{7}{c}{TIP4P/2005} \\ \hline
        -96.9 & 1.24 & 0.953 & 6855 & 59.9 & 93.6 & -39.1 \\
        -87.6 & 1.38 & 0.956 & 6796 & 60.3 & 116.8 & -49.2 \\
        -77.0 & 1.59 & 0.960 & 23814 & 61.2 & 158.5 & -67.3 \\
        -73.0 & 1.68 & 0.961 & 11385 & 61.5 & 178.2 & -75.8 \\ 
        -61.9 & 2.01 & 0.965 & 11006 & 62.0 & 255.3 & -109.3 \\
        -54.9 & 2.29 & 0.968 & 10570 & 62.8 & 337.3 & -144.9
    \end{tabular}
    \caption{NVT seeding data for the DPMD and TIP4P/2005 models, including the nucleation pressure ($P$), the critical radius ($R_c$), the liquid density ($\rho _l$), the total number of water molecules in the system ($N_T$), the surface tension ($\gamma$), the free energy barrier ($\Delta G^*$), and the logarithm of the nucleation rate (log$_{10}J$).}
    \label{tab_seeding}
\end{table*}

\subsection{Determination of the Tolman length}

Another quantity we can extract from our simulations is the Tolman length, which describes the deviation of the surface tension with respect to its value at the planar interface and, therefore, the coexistence conditions. The Tolman length can also be defined as the deviation of the surface of tension from the equimolar dividing surface. In 1949, Tolman showed that the change in surface tension with curvature follows the equation \cite{tolman1949effect}

\begin{equation}
    \gamma=\frac{\gamma _0}{1+\frac{2\delta}{R_c}}
    \label{tolman}
\end{equation}
where $\gamma _0$ is the value of the surface tension under coexistence conditions and $\delta$ is the Tolman length. This quantity has been extensively studied for Lennard-Jones particles \cite{kashchiev2000nucleation,van2009direct,bykov1999patching,lei2005tolman,julin2010thermodynamically,kashchiev2020nucleation}, Hard Spheres \cite{montero2020interfacial} and other systems \cite{van2009direct,troster2011positive,iwamatsu1997temperature,bykov1999patching}. Here we make use of Tolman's expression and compute $\delta$ for both the DPMD and TIP4P/2005 models at 296.4 K. We fit our surface tension and critical radius data (obtained from the NVT seeding simulations) to Eq. \ref{tolman}, performing a non-linear regression. We choose to have $\gamma _0$ and $\delta$ as fitting parameters, despite having estimated the former from the DC simulations. We took this approach in order to corroborate the value of $\gamma$ obtained from both approaches. \\

\begin{figure}
    \centering
    \includegraphics[width=\linewidth]{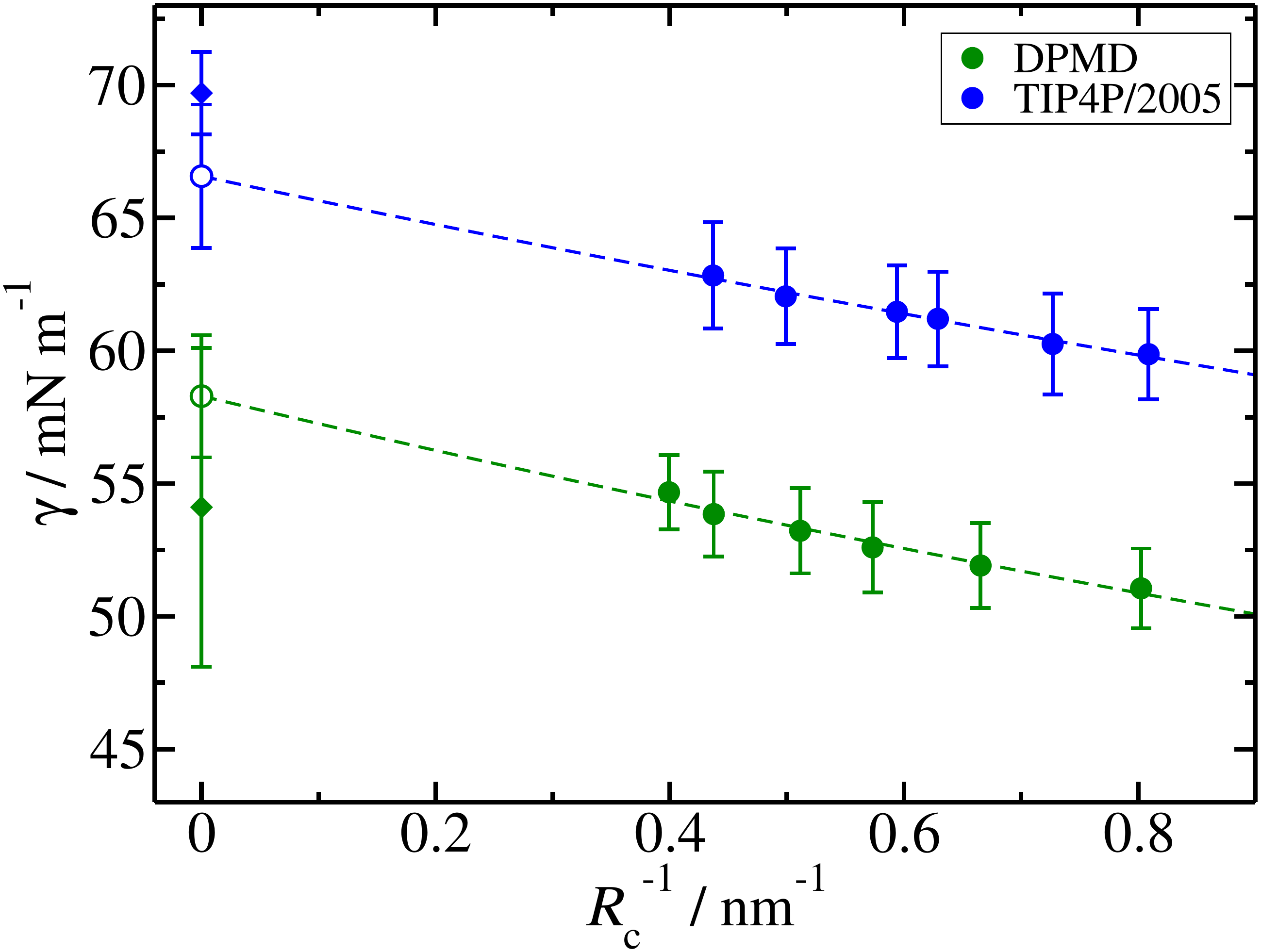}
    \caption{Surface tension $\gamma$ against the inverse of the critical radius ($R_c^{-1}$). Filled points represent the surface tension obtained from Laplace's equation (Eq. \ref{eq3}). The dashed lines indicate the change in surface tension according to Tolman's equation (Eq. \ref{tolman}), where the parameters $\delta$ and $\gamma _0$ were obtained through non-linear regression. We plot the values of surface tension at planar interface ($R_c^{-1}$=0) obtained from the fit with empty points. The surface tension values obtained from DC are also represented here with filled diamonds, also at $P=0$.}
    \label{figure4}
\end{figure}

In Figure \ref{figure4} we plot the surface tension against the inverse of the critical radius for the DPMD (green filled points) and TIP4P/2005 (blue filled points) models. From non-linear regression we obtain $\delta=(0.091\pm 0.008)$ nm for the DPMD model and $\delta=(0.070\pm 0.004)$ nm for the TIP4P/2005 model. Both values are positive as expected, since the surface tension decreases for smaller bubbles. In Figure \ref{figure4} we also show with dashed lines how the surface tension changes against the inverse of the critical radius according to Eq. \ref{tolman}. From the fit we also obtain values for the surface tension at planar interface of 58 $\pm$ 2 and 67 $\pm$ 3 mN·m$^{-1}$ for the DPMD and TIP4P/2005 models respectively. These values are in agreement, within statistical uncertainties, with those obtained from DC simulations (54 and 70 \cite{vega2007surface} mN·m$^{-1}$ for DPMD and TIP4P/2005 respectively). The uncertainty of $\delta$ is simply determined by the error in the non-linear fit, while the uncertainty in $\gamma _0$ is the sum of the errors coming from NVT seeding simulations (see Section \ref{cntsec}) and the error in the fit to Eq. \ref{tolman}. \\

Our results are also in good agreement with previous simulation results \cite{magaletti2021water,min2019bubbles} (at different temperatures: 0.199 nm at 300 K \cite{min2019bubbles}, 0.09 nm at 250 K \cite{magaletti2021water} and 0.18 nm at 350 K) which indicate that the Tolman length is positive for water bubbles, in contrast to the negative sign in water droplets (i.e. condensation) \cite{joswiak2016energetic,joswiak2013size}, where the surface tension increases with curvature. We can establish direct comparison with the calculations from Menzl \emph{et al.} \cite{menzl2016molecular}, which were performed with TIP4P/2005 at the same temperature (296.4 K). They obtained values of $\delta=0.195$nm and $\gamma _0$=82.79 mN·m$^{-1}$, which while having the same order of magnitude, moderately disagree with our calculations of $\gamma _0$ and $\delta$. In the case of Ref. \cite{menzl2016molecular} the employed local order parameter does not necessarily identify an accurate value of the radius, but is instead used to bias the sampling of the configurational space for US simulations, and may therefore not represent accurate values of $\delta$ and $\gamma _0$. As mentioned before, the good agreement in the nucleation free energy barrier between US and seeding calculations is a good sign. In our case, a good indicator for the calculation of the Tolman length, although not definitive, is the fact that the surface tension obtained from the non-linear regression to Eq. \ref{tolman} provides a value of $\gamma _0$ that matches within the uncertainty the surface tension obtained from DC simulations, as aforementioned.

\subsection{Liquid-vapor interfacial characterization}

Finally, we examined the organization of the liquid-vapor interface in our simulations for both planar and curved interfaces. The water-air planar interface has been widely studied in the past \cite{tang2020molecular} with IR vibrational spectra experiments \cite{nagata2015surface,bonn2015molecular,nihonyanagi2013structure,nihonyanagi2011unified}, \emph{ab initio} \cite{liang2019ab,ohto2015toward,kessler2015structure,vassilev2001ab} and MD simulations \cite{khatib2016molecular,lee1984structure,fan2009structure}, all of which generally agree regarding the orientation of water molecules near the interface. Interfacial molecules closer to the vapor phase tend to orient the O-H bond parallel to the surface normal vector and expose the H atom, resembling the (1000) crystal face of ice Ih (only in the dimension perpendicular to the surface) \cite{fan2009structure}. The (1000) ice Ih-liquid is the direction with lower interfacial free energy of the different solid-liquid interfaces \cite{espinosa2016ice}, therefore in the liquid-vapor interface a similar arrangement may also reduce the surface tension, apart from maximizing the enthalpic gain of the more exposed interfacial molecules to the vapor phase. To perform this analysis, we follow the same approach as in Refs. \cite{fan2009structure, vassilev2001ab} where, once the interfacial region has been identified, the angle formed by the O-H bonds and the vector normal to the interface ($\theta _H$) is computed. Fan \emph{et al.} \cite{fan2009structure} and Vassilev \emph{et al.} \cite{vassilev2001ab} identified two distinct layers in non-curved interfaces: an external layer, which comprehends the region in which the density is between 5-50\% of the bulk liquid, and an internal layer, where the density ranges between 50-95\% of the bulk liquid density. Fan \emph{et al.} \cite{fan2009structure} measured the orientation distributions for 
the TIP3P, TIP4P-EW, TIP5P and SPC/E water models at 300K, and for comparison purposes, we perform the same analysis with the DPMD and TIP4P/2005 models, and obtain results also for curved interfaces. \\

\begin{figure*}
    \centering
    \begin{tabular}{c c|c|c}
         \multicolumn{2}{c}{(a)} & \multicolumn{2}{c}{(b) \hspace{0.28\linewidth} (c)} \\
         \\
         & & TIP4P/2005 & DPMD \\
        \rotatebox{90}{\ \ \ Planar interface} &

        \includegraphics[width=0.25\linewidth]{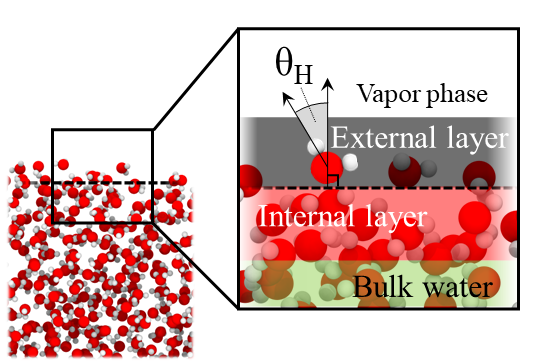} & \includegraphics[width=0.3\linewidth]{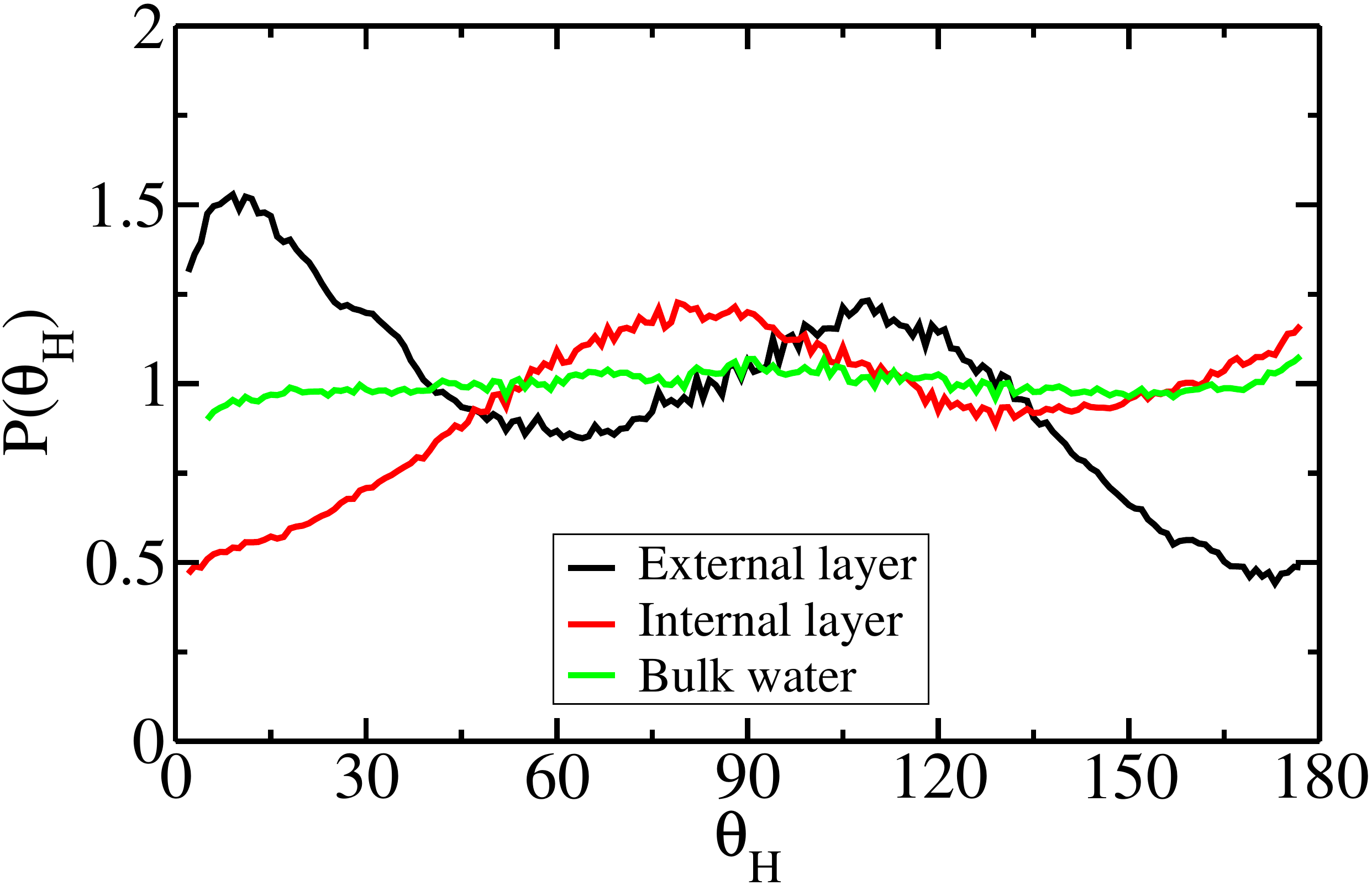} & \includegraphics[width=0.3\linewidth]{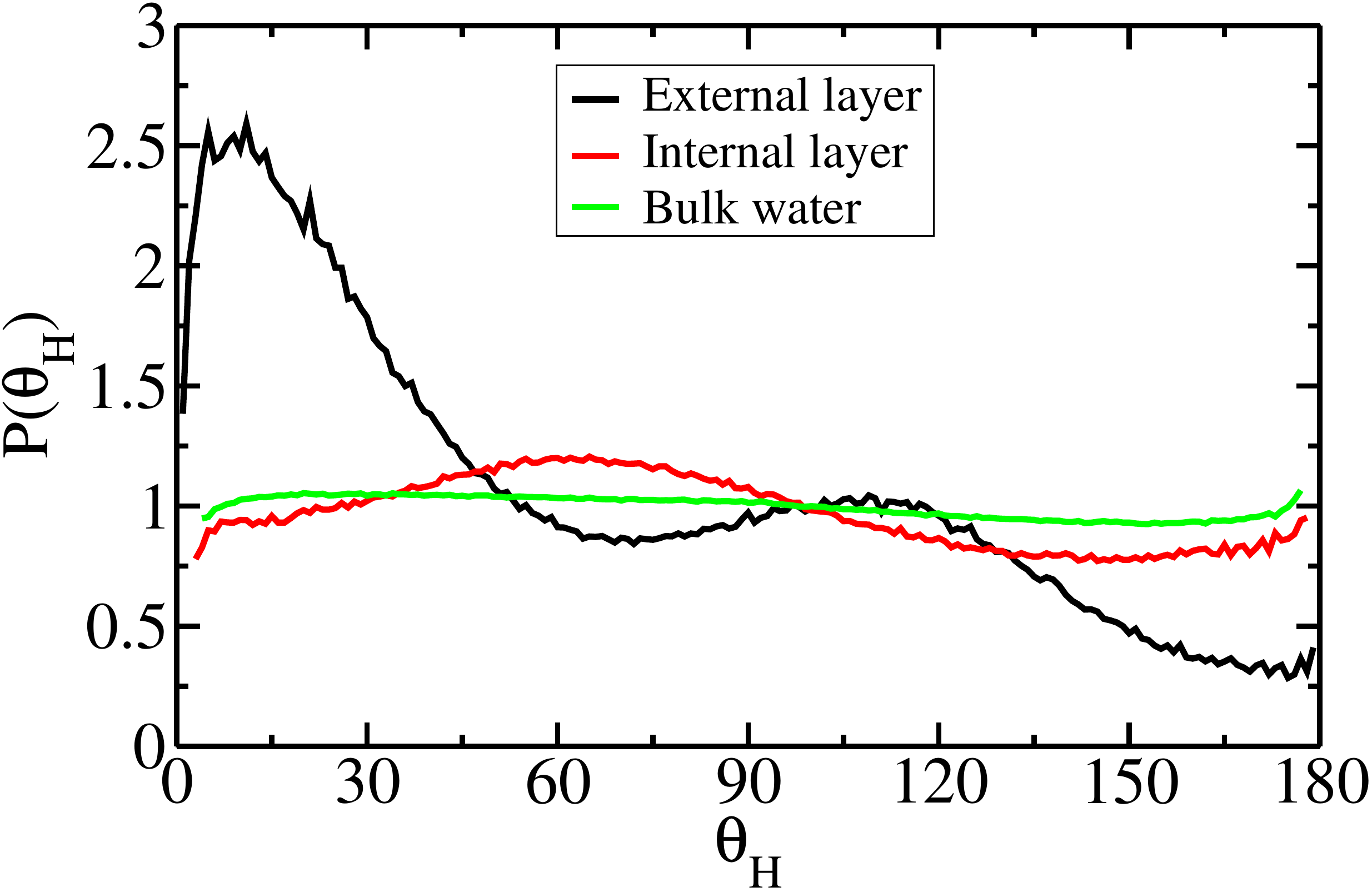} \\  \hline
        %{histo_ang_tip4pplan.pdf} & \includegraphics[width=0.3\linewidth]{histo_angdpmdplano.pdf} \\  \hline
 & & & 
\\

        \rotatebox{90}{\ \ \ \ Curved interface} &
        \includegraphics[width=0.25\linewidth]{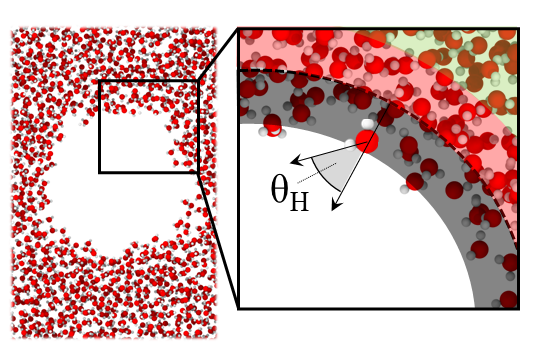} &
        \includegraphics[width=0.3\linewidth]{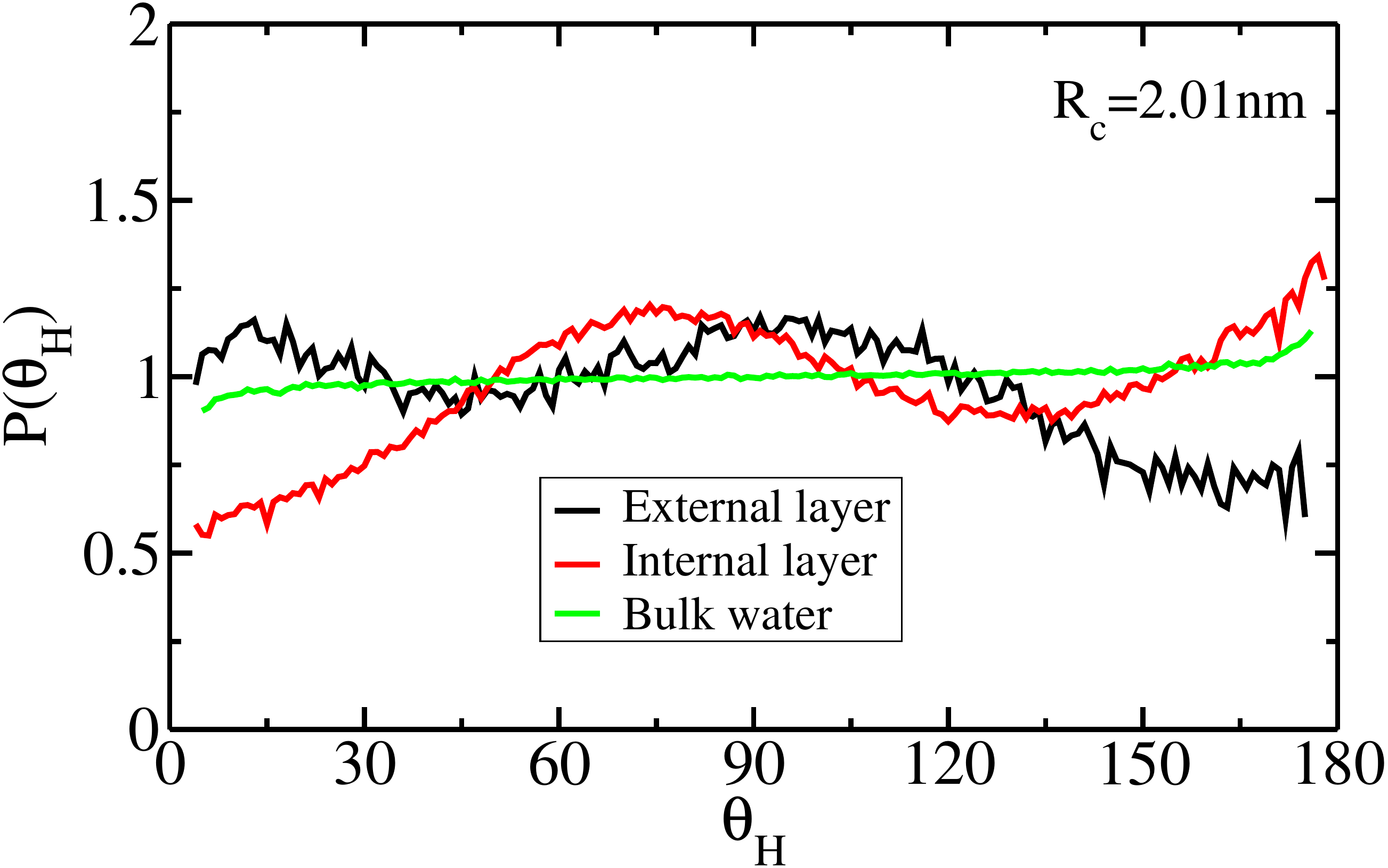} & \vspace{1in} \includegraphics[width=0.3\linewidth]{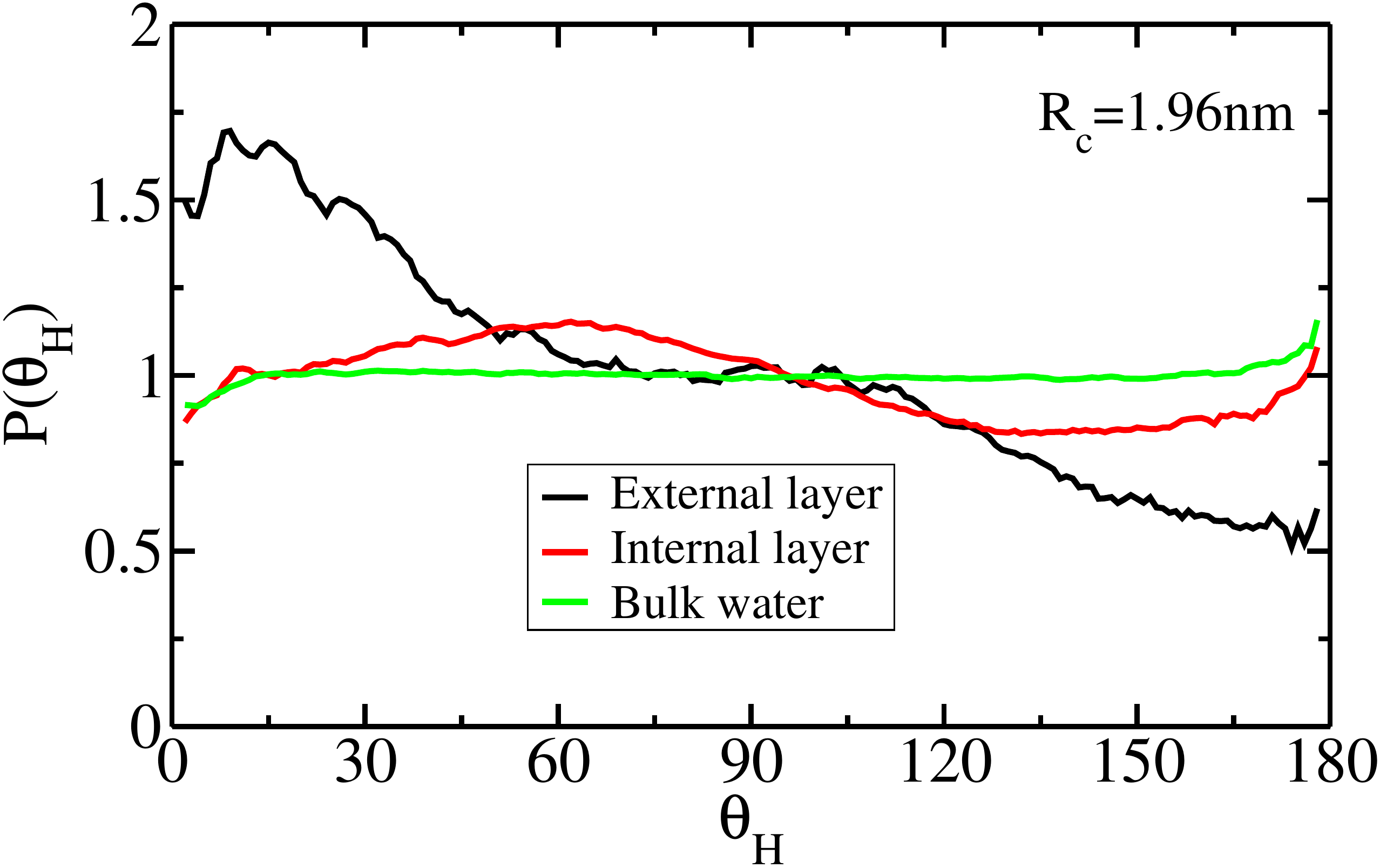} %\includegraphics[width=0.3\linewidth]{dists.pdf} & \vspace{1in} \includegraphics[width=0.3\linewidth]{hisdc.pdf}
    \end{tabular}
    \vspace{-2.3cm}
    \caption{(a) Top: Slab snapshot (including only a reduced amount of molecules for better visualization) of a planar interface. Bottom: Snapshot for a curved interface. (b) Normalized histograms of $\theta_H$ at the external region of the interface, the internal one, and bulk water for the TIP4P/2005 model for planar (top) and curved (bottom) interfaces. (c) Same as in (b) but for the DPMD model.}
    \label{figure5}
\end{figure*}

In Figure \ref{figure5}(a) we depict how the different interfacial regions and $\theta _H$ angle are identified for both planar (top) and curved (bottom) interfaces.  There, the dashed black line indicates the liquid-vapor interface as defined by our order parameter (see section \ref{cnt} and Supplementary Material (SM)). In the zoomed image, the internal and external layers of the interface are labelled and highlighted in black and red, as well as the bulk liquid and vapor phases in green and white respectively. The two arrows indicate the normal vector to the interface and the O-H bond vector, and the angle between these two vectors defines $\theta_H$. In Figure \ref{figure5}(b-c) We plot the $\theta_H$ distributions normalized by the random distribution sin($\theta _H$). Consistent with previous calculations \cite{fan2009structure,lee1984structure,vassilev2001ab,kessler2015structure}, the molecules expose one hydrogen atom to the vapor phase in the external layer, as can be seen in Figure \ref{figure5}(b)(top) for the TIP4P/2005 model, where we show the probability distribution of the angle $\theta _H$ for the different interfacial regions. Smaller angles are more probable in the external layer than in the bulk, where all directions are equally probable (flat green line). As discussed in Ref. \cite{fan2009structure}, the internal layer of the interface also displays preferential orientations, in order to maximise the interactions with the structure created in the external layer, in an arrangement that leads to a maximum near $\theta _H\sim$80º. It must be noted that the preferential molecular orientations are not fixed, but rather transient. The distributions presented in Figure \ref{figure5}(b)(top) are also consistent with the results by Fan \emph{et al.} \cite{fan2009structure} for other rigid semi-empirical models. \\

We extended this analysis to curved interfaces, using our NVT seeding simulations. In curved interfaces, the vector normal to the interface points at the bubble center. In Figure \ref{figure5}(b)(bottom) we show the probability distribution of $\theta _H$ for the three interfacial regions, and observe how these resemble those of planar interface. A significant loss of ordering is observed for the curved interface since the peaks at $\sim$ 15º and $\sim$ 115º in the external layer, and at $\sim$ 80º in the internal layer are less prominent compared to the the corresponding peaks for the flat interface. This could result of the curved geometry of the interface sterically impeding a better rearrangement of the interfacial molecules. \\

We also evaluated the $\theta _H$ distribution for the DPMD model. In Figure \ref{figure5}(c)(top) we present the distribution of the three interfacial regions for the planar interface at 296.4 K. There is an obvious preferential orientation towards small angles in the external layer, once again indicating the preference of molecules to expose one hydrogen atom to the vapor phase. The distribution is remarkably more prominent for the DPMD model relative to TIP4P/2005, although both models result in similar molecular arrangements. In Ref. \cite{vassilev2001ab}, this same structure was found in \emph{ab initio} calculations using PW91 functional, confirming that the DPMD model based on the SCAN functional yields similar results to those obtained from \emph{ab initio} MD simulations using different density functionals. \\

Looking at the DPMD model distributions for curved interfaces (Figure \ref{figure5}(c)(bottom)), we still find a preferential ordering towards small angles, although it is much less pronounced than in the case of planar interfaces, in agreement with TIP4P/2005 simulations. The bubbles generated by both models shown here (TIP4P/2005 and DPMD) have comparable sizes, however the effect of curvature is more dramatic in the case of the DPMD model. Other bubble size distributions are reported in the SM, where we observe that there is a slight change in the distributions, with bigger bubbles resulting curves more similar to those found at coexistence (Figure S3).
%We have established that interfacial water molecules also preferentially orient a O-H bond towards the curved bubble interfaces. Nonetheless, steric effects may be causing cause less prominent organization in this interfacial distribution for bubbles between 1-2 nm in size. \\

\section{Conclusions}

In this work, we explored the liquid-vapor phase behavior of a Deep Potential model based on \emph{ab initio} energies and forces. This model was derived from the SCAN approximation of density functional theory \cite{zhang2021phase}. The model has been shown to reproduce the phase diagram for the different ice phases \cite{zhang2021phase} and to have a liquid-liquid phase transition in the supercooled regime \cite{tom2020signatureLLT}. We computed the phase diagram via DC simulations, and adjusted our vapor equilibrium densities to the experimental values, resulting in a shift of 40 K in the DPMD model. Once the model was tested and shifted, we compare its equilibrium properties with experimental data and the TIP4P/2005 model \cite{abascal2005general}, one of the most benchmarked classical models for water \cite{menzl2016molecular,khatib2016molecular,vega2006vapor,vega2007surface}. The surface tension of the DPMD model is lower than the one obtained from TIP4P/2005 and experiments. Nonetheless, by construction the DPMD model provides accurate results for other properties such as the vapor saturation pressure for a wide range of temperatures, between 300 and 600 K (Figure \ref{figure1}). Overall, once the temperature is shifted, the model reproduces most of the properties of liquid-vapor equilibrium, despite not having this regime included in its training. This result is especially remarkable since this is the first Deep Potential-based model to provide liquid-vapor properties in a computationally affordable time, with the primary drawback being the $\sim$20\% deviation in the surface tension.
%We note that the temperature shift was applied in order to obtain a better match of the vapor phase behaviour, nonetheless this shift affects negatively on the surface tension prediction. With no temperature shift, the surface tension of the DPMD model matches the experimental one at temperatures above 450 K, and only underestimates it by $\sim 5 \%$ at T $<$ 450 K.  \\

Moreover, we study bubble nucleation in the cavitation regime. We make use of the NVT seeding method, a rare event technique already shown to be successful in determining the nucleation free energy barrier and the nucleation rate for simpler systems \cite{rosales2020seeding,sanchez2020equivalence} in cavitation events. 
%Using this technique, properties of a critical bubble are measured for long times in the canonical ensemble. Since the seeding requires a careful choice of the local order parameter to determine the critical bubble size, we make use of the equidensity criterion, i.e. the critical radius is found at the interfacial region in which the density is the average between the liquid and vapor densities. 
We first performed NVT seeding calculations at 296.4 K, a temperature at which previous data from Umbrella Sampling are available for the TIP4P/2005 model and we confirm that the seeding technique provides consistent results with those from Menzl \emph{et al.} \cite{menzl2016molecular}. The DPMD water model provides higher nucleation rates than the TIP4P/2005 model under the same stretching conditions. We show that this quantitative difference can be explained by the difference in surface tension between models, which persists for curved interfaces (Figure \ref{figure3}). Our results highlight once more the relevance of the surface tension and its change with curvature to critically control nucleation events \cite{magaletti2021water,menzl2016molecular}. We could have obtained closer agreement between the TIP4P/2005 nucleation rates with those from the DPMD model if we did not apply the temperature shift, nonetheless we prioritised adjusting the model to obtain better equilibrium densities, in line with prior studies of water properties using the SCAN-derived DPMD model \cite{yaoyi2020scanshift, tom2020signatureLLT, piaggi2021phaseequilibrium}. \\

Furthermore, we provide an estimate of the Tolman length by performing a non-linear regression to the relevant expression \cite{tolman1949effect}, which describes how the surface tension changes with curvature (Eq. \ref{tolman}). Using data from our NVT seeding simulations we obtain estimates of $\delta=(0.091\pm 0.008)$ nm for the DPMD model and $\delta=(0.070\pm 0.004)$ nm for the TIP4P/2005 model, both at 296.4 K. This confirms that a Deep Potential-based model also predicts a positive sign of the Tolman length, confirming previous results showing a decrease of the surface tension with curvature for the case of water bubbles \cite{magaletti2021water,min2019bubbles,menzl2016molecular}. \\

Finally, we studied the orientation of the water molecules in the interface, corroborating previous studies that have indicated that the molecules closer to the vapor phase have a preference so as to expose an hydrogen atom facing the vapor \cite{tang2020molecular,nagata2015surface,bonn2015molecular,nihonyanagi2013structure,nihonyanagi2011unified,liang2019ab,ohto2015toward,khatib2016molecular,lee1984structure,fan2009structure}. We quantify this behavior by measuring the angle formed between the normal vector to the interface and the O-H bond. We find that a preferential molecular orientation appears for both the TIP4P/2005 and DPMD models. We also confirm that this phenomenon also takes place in the curved interface of water bubbles, although the possibility to orient more O-H bonds towards the vapor is diminished for curved interfaces due to the increasing curvature of the bubbles. \\

Overall, this study confirms that machine-learning \emph{ab initio} based models that capture more molecular details than semi-empirical models are viable for prediction of equilibrium as well as dynamic properties that require large system sizes and long sampling times. The computational cost of \emph{ab initio} based models in long scale Molecular Dynamics is now affordable being only an order of magnitude greater than that for empirical potentials, thanks to recent improvements such as the compressed Deep Potential modelling scheme \cite{lu2022dp}. Further work and models trained with more liquid and vapor data will only improve the already existing models, which will gradually be better in describing the real behaviour of water. \\

\section*{Acknowledgments}

I.~S.-B. acknowledges funding from Derek Brewer scholarship of Emmanuel College and EPSRC Doctoral Training Programme studentship, number EP/T517847/1. J.~R.~E.  acknowledges funding from the Roger Ekins Research Fellowship of Emmanuel College, Oppenheimer Research Fellowship of the University of Cambridge and the Ramon y Cajal fellowship (RYC2021-030937-I). M.~C.~M and A.~Z.~P acknowledge the “Chemistry in Solution and at Interfaces” (CSI) Center funded by the U.S. Department of Energy Award DE-SC$001934$ and Award DE-SC0002128. Computational resources were provided by the Princeton Research Computing at Princeton University which is a consortium of groups led by the Princeton Institute for Computational Science and Engineering (PICSciE) and the Office of Information Technology's Research Computing.

\bibliographystyle{ieeetr}

\clearpage

\counterwithin{equation}{section}
\counterwithin{table}{section}
\setcounter{section}{0}
\setcounter{figure}{0}
\renewcommand\thesection{S\Roman{section}}   
\renewcommand\thefigure{S\arabic{figure}} 
\renewcommand\thetable{S\arabic{table}} 
\renewcommand\theequation{S\arabic{equation}}    
\onecolumngrid
\setcounter{page}{1}
\begin{center}
    {\large \textbf{Supplementary Material: A Deep Potential model for liquid-vapor equilibrium and cavitation rates of water} \\ \par} \vspace{0.3cm}
    Ignacio Sanchez-Burgos$^{1,2}$, Maria Carolina Muniz$^{2}$, Jorge R. Espinosa$^{1,3}$ and Athanassios Z. Panagiotopoulos$^{2,*}$ \\ \vspace{0.15cm}
    \emph{$[1]$ Maxwell Centre, Cavendish Laboratory, Department of Physics, \\ University of Cambridge, J J Thomson Avenue, Cambridge CB3 0HE, United Kingdom. \\
    $[2]$ Department of Chemical and Biological Engineering, \\ Princeton University, Princeton, New Jersey 08544, USA. \\
    $[3]$ Departamento de Química Física, Facultad de Ciencias Químicas, \\ Universidad Complutense de Madrid, 28040 Madrid, Spain. \\
    * = To whom correspondence should be sent. email: azp@princeton.edu} \\

\end{center}
\thispagestyle{empty}

\section{Equation of state}

\begin{figure}[h]
    \centering
    \includegraphics[width=0.5\linewidth]{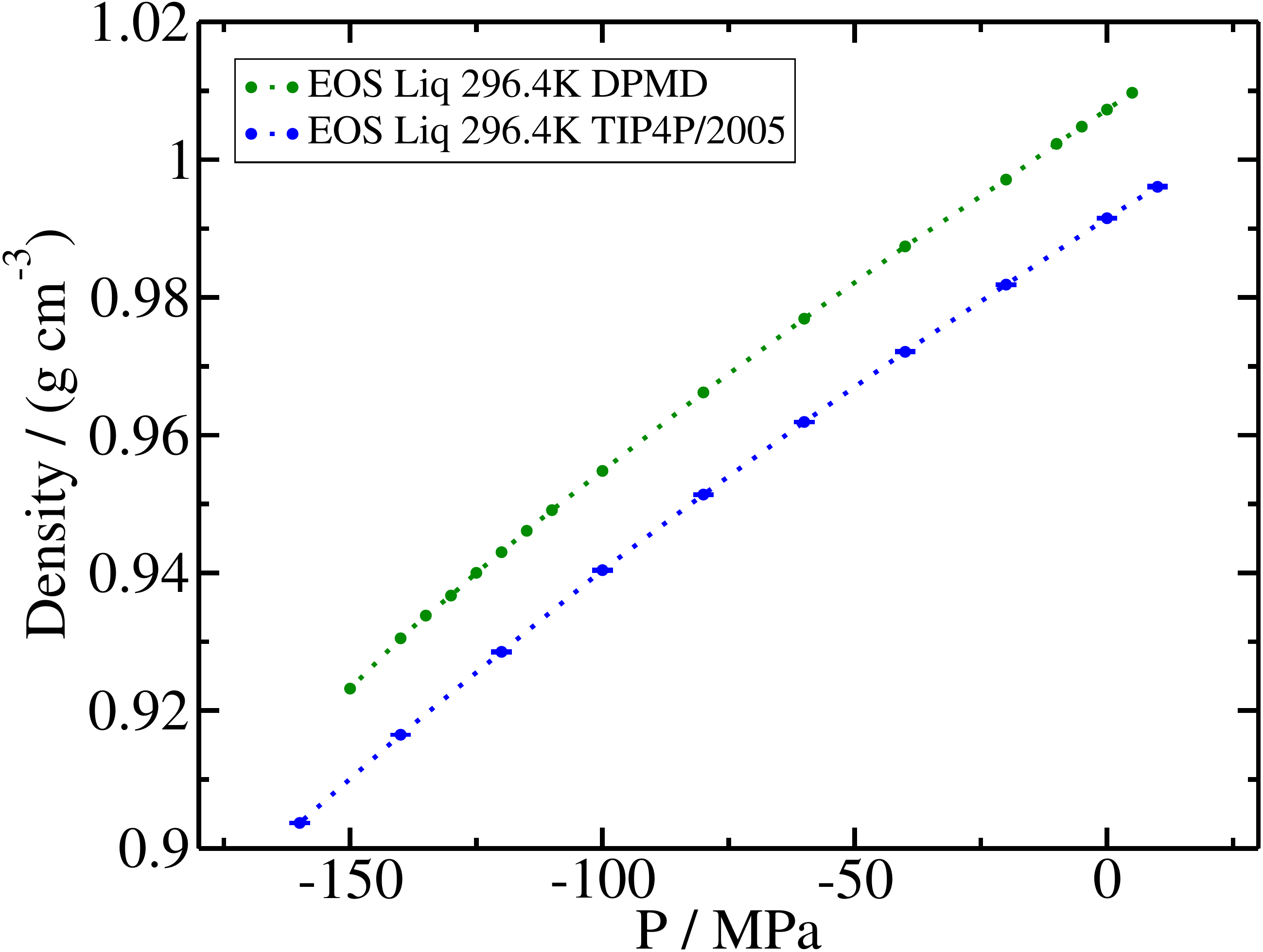}
    \caption{Density against pressure equation of state at 296.4 K for the DPMD and TIP4P/2005 models.}
    \label{eosliq}
\end{figure}

As mentioned in the main text, we corroborate that we can obtain the density of the liquid phase surrounding the critical bubbles. Then, with such density, by means of the $\rho _l$ against $P$ equation of state we obtain the pressure of the liquid phase, which matches the pressure obtained through the virial expression. In Figure \ref{eosliq} we show the equation of state at 296.4 K for the DPMD and TIP4P/2005 models.

\section{Bubble sphericity}
\label{sec1}

\begin{figure}[h]
    \centering
    \begin{tabular}{cc}
         (a) & (b) \\
         \includegraphics[width=0.2\linewidth]{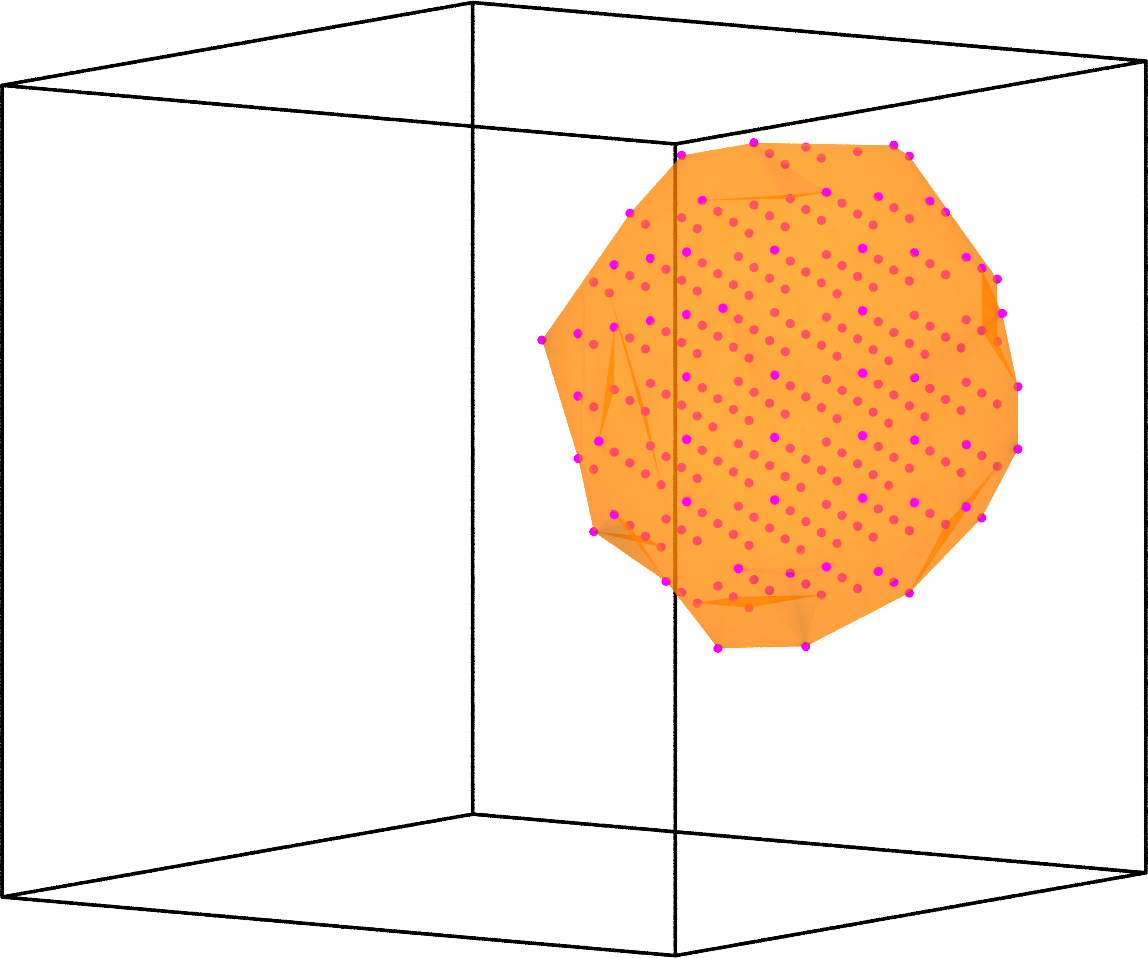} & \begin{tabular}{ccc}
             \includegraphics[width=0.23\linewidth]{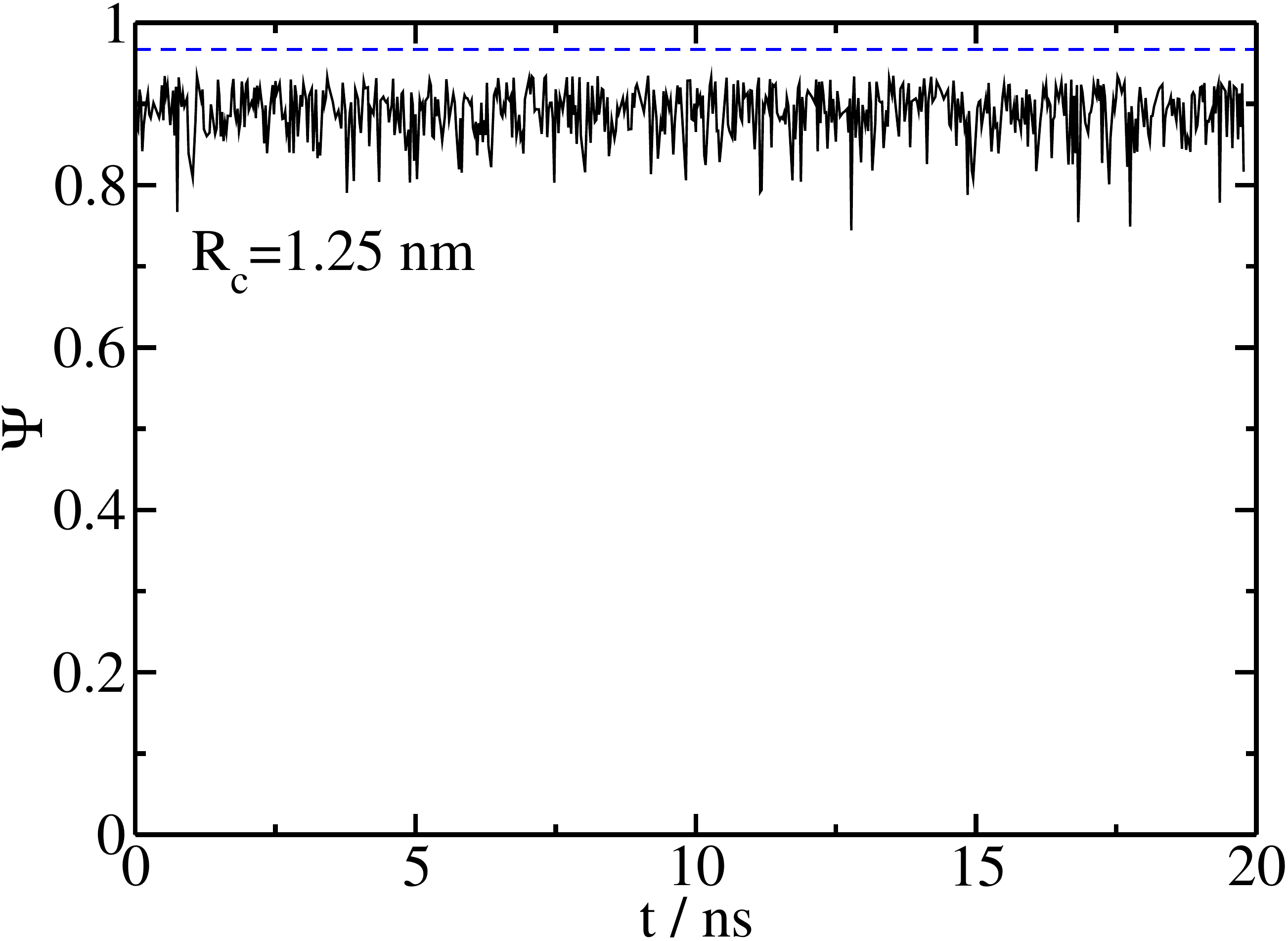} & \includegraphics[width=0.23\linewidth]{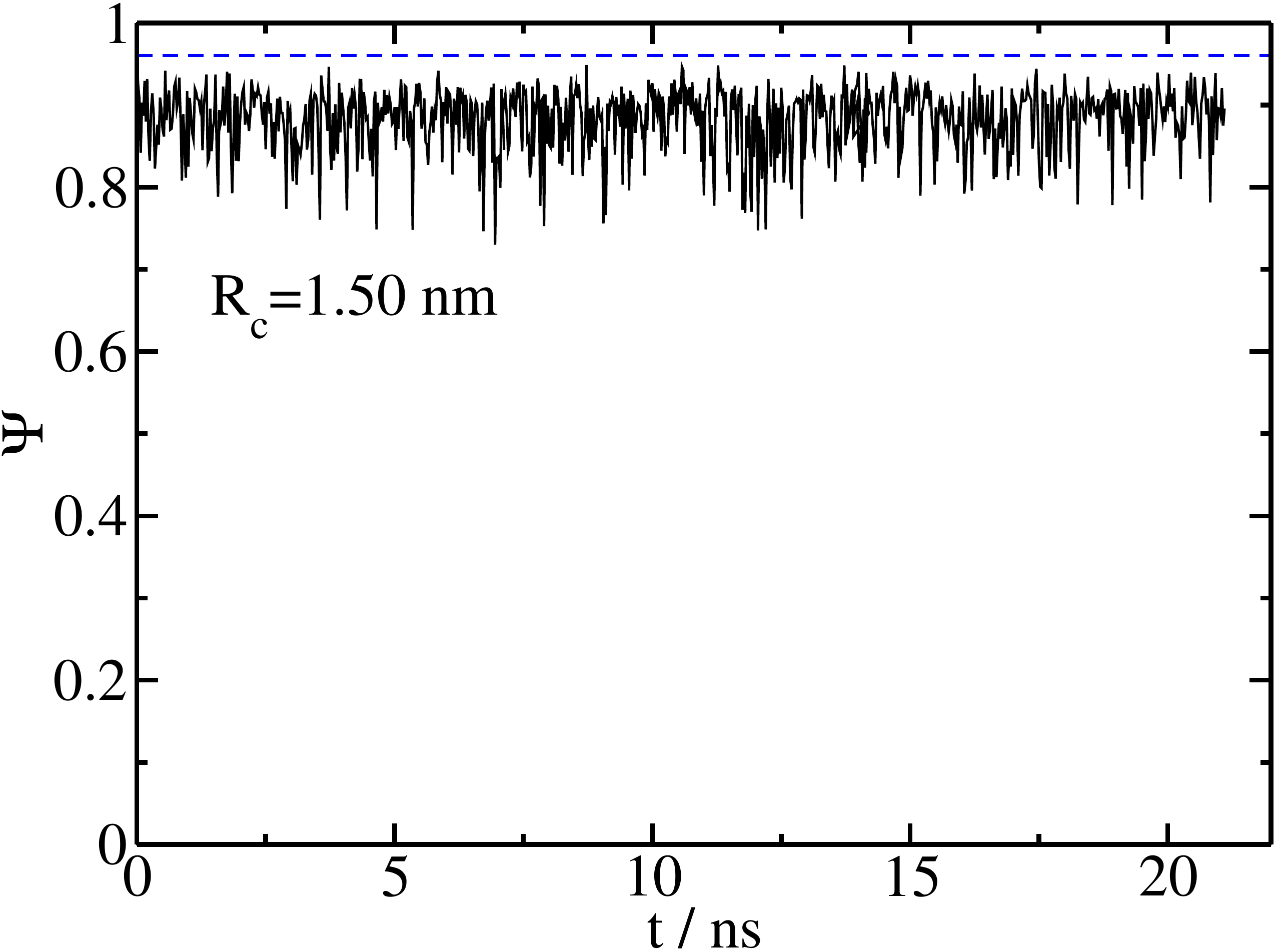} & \includegraphics[width=0.23\linewidth]{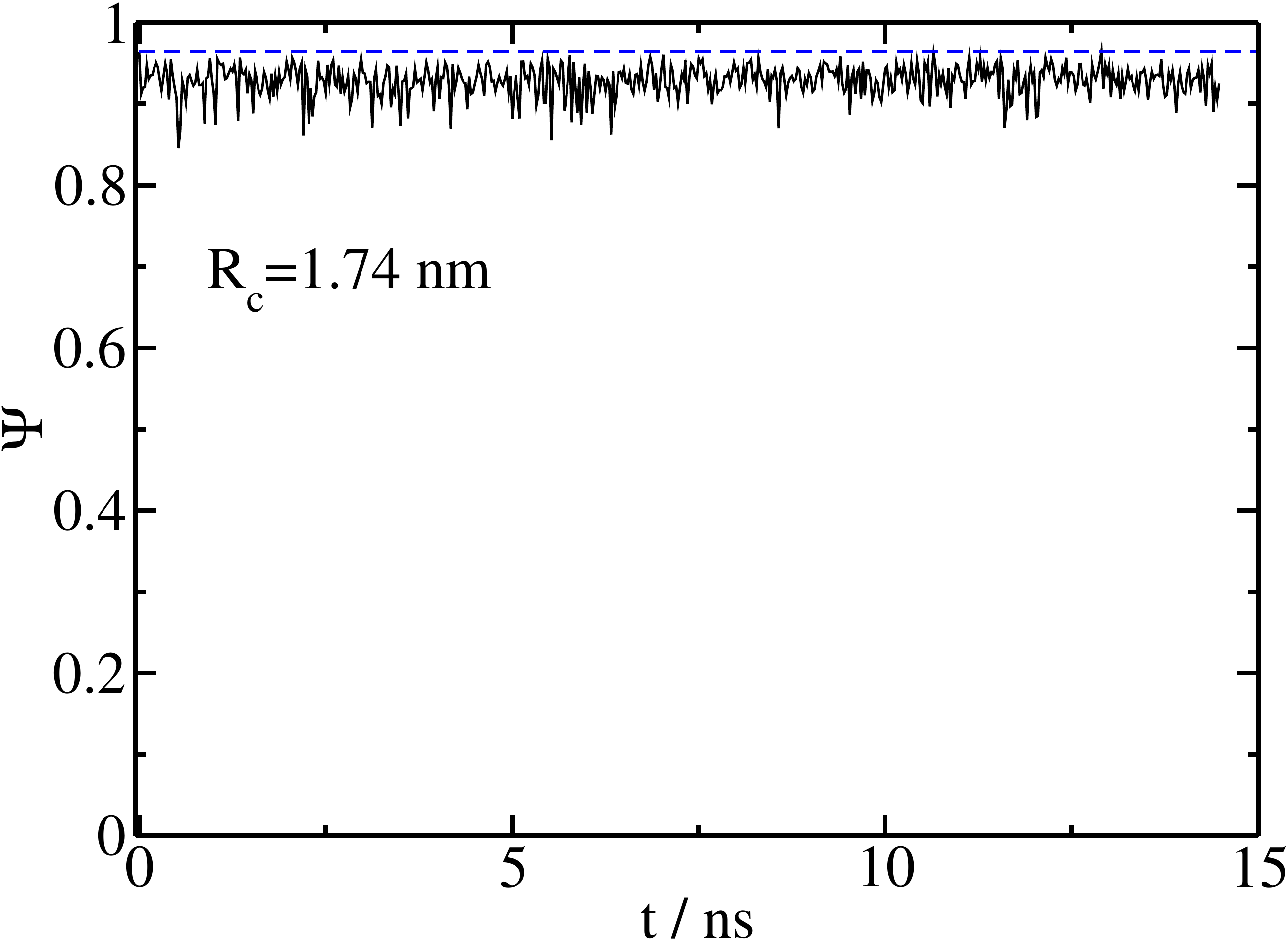} \\
             \includegraphics[width=0.23\linewidth]{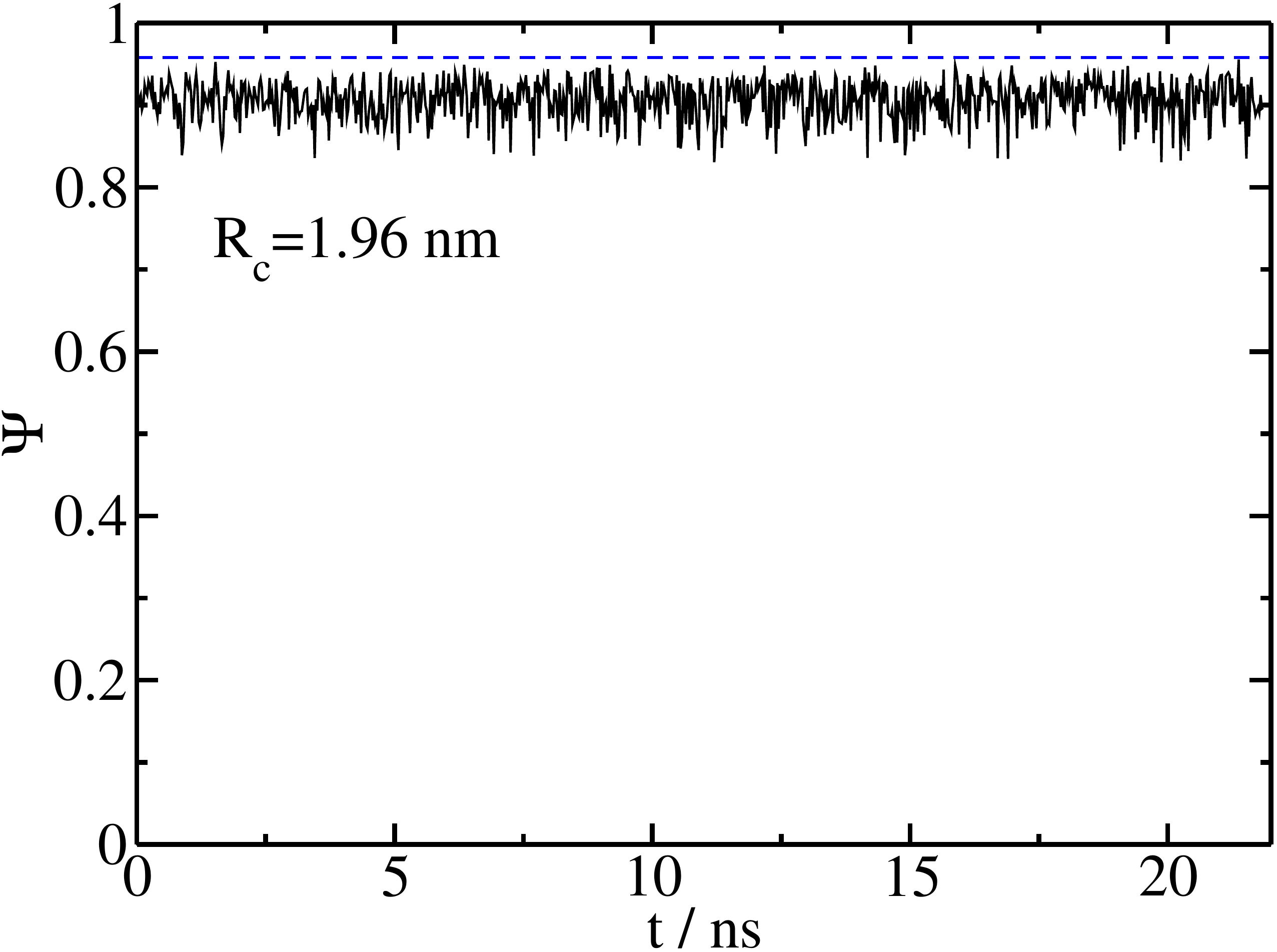} & \includegraphics[width=0.23\linewidth]{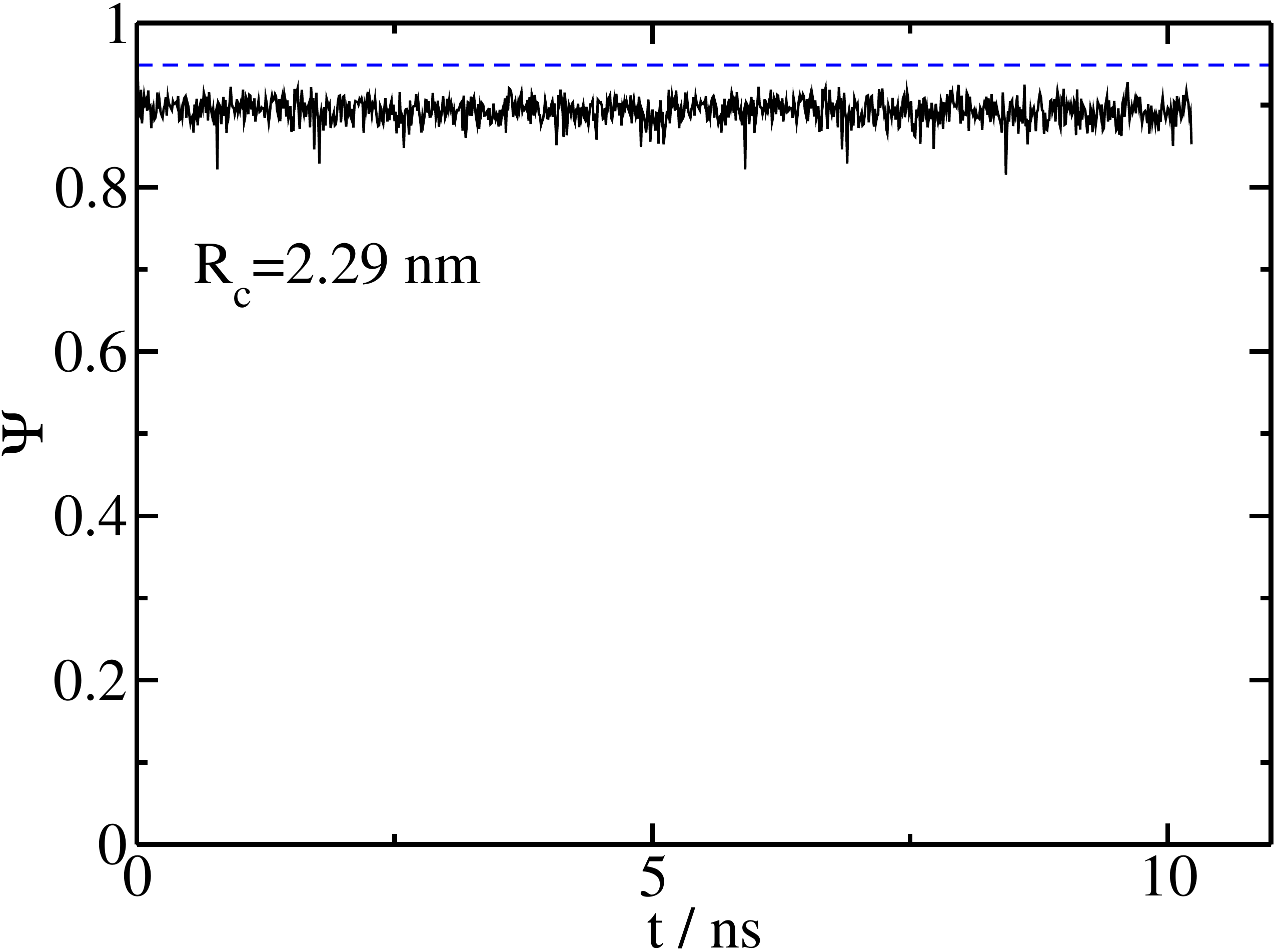} & \includegraphics[width=0.23\linewidth]{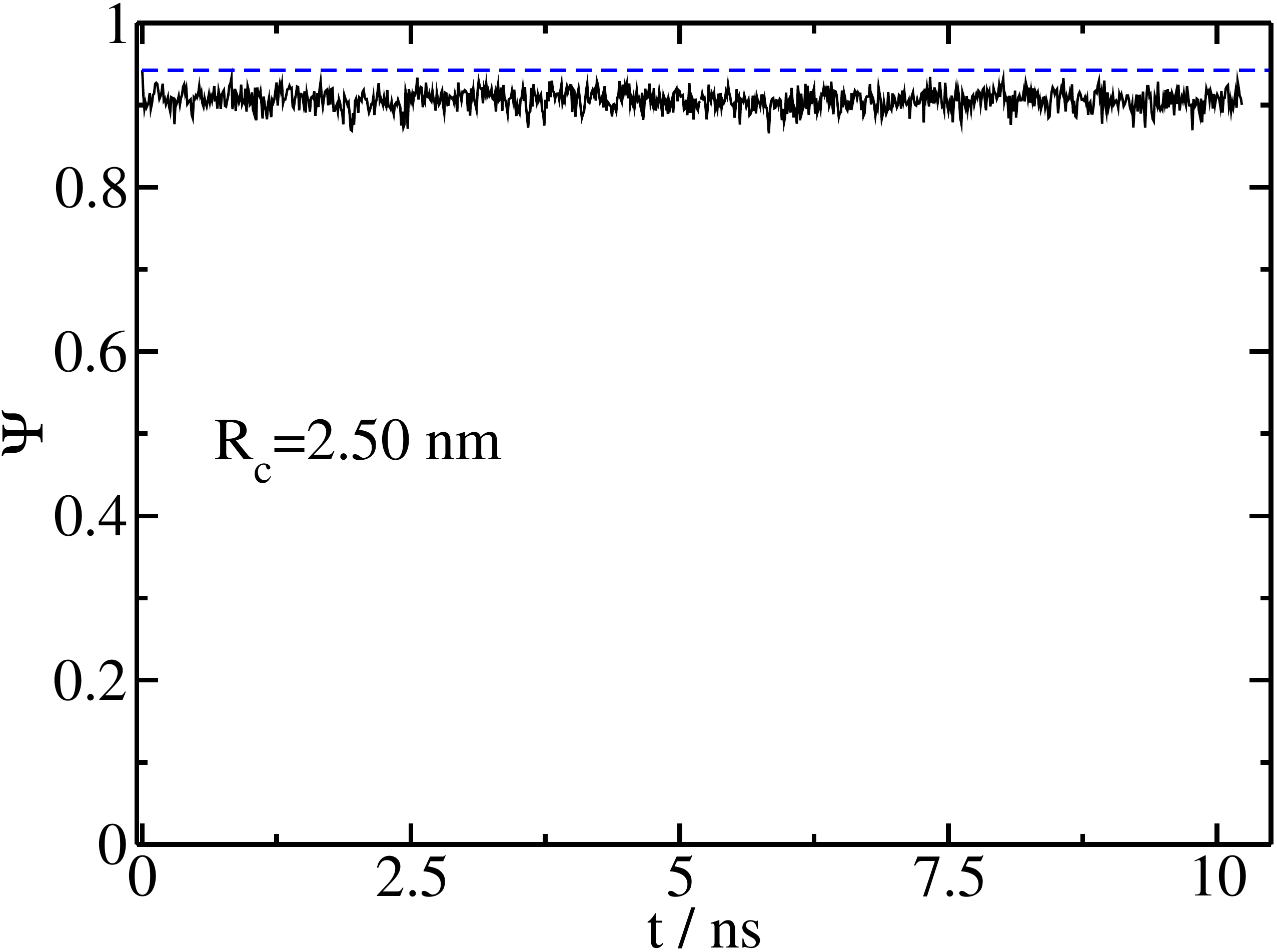}
         \end{tabular}
    \end{tabular}
    \caption{(a) Snapshot of a vapor cluster (R$_c$=2.29 nm) identified as detailed in section \ref{sec1}. The identified surface is colored in orange. (b) $\Psi$ against time for the 6 simulated bubbles in Seeding simulations with the DPMD model.}
    \label{sphe}
\end{figure}

In order to quantify how spherical the simulated bubbles are, we take the following approach: We divide the simulation box into smaller cells of $\sim 200$\r{A}$^3$ each. We classify the cells as liquid or vapor based on their local density, taking 0.3 g cm$^{-3}$ as threshold density. We then identify the surface available to the vapor cluster by means of a Surface mesh algorithm \cite{edelsbrunner1994threeKK,stukowski2014computationalKK} available with the OVITO software package \cite{stukowski2009visualizationKK}. From this, we obtain the surface and volume that corresponds to the vapor cluster. Using his information we compute the cluster sphericity ($\Psi$) as \cite{wadell1935volumeKK}:

\begin{equation}
    \Psi=\frac{\pi^{1/3}(6V)^{2/3}}{S}
\end{equation}

where V is the cluster volume and S the surface. $\Psi$ is equal to 1 for a perfect sphere and decays for less spherical geometric bodies. In Figure \ref{sphe} we show a rendered image of the surface created for a vapor cluster (depicted in orange), as well as $\Psi$ against time for our Seeding simulations with the DPMD model. The values of $\Psi$ exceed 0.85 in all cases which means a high sphericity of the simulated bubbles. We show with a blue line $\Psi$ for the initial configuration in which a perfect sphere has been generated, and represents the maximum sphericity that our parameter is able to account for.

\section{Orientational analysis}

\begin{figure}[h]
    \centering
    (a) \hspace{0.42\linewidth} (b) \\ \includegraphics[width=0.45\linewidth]{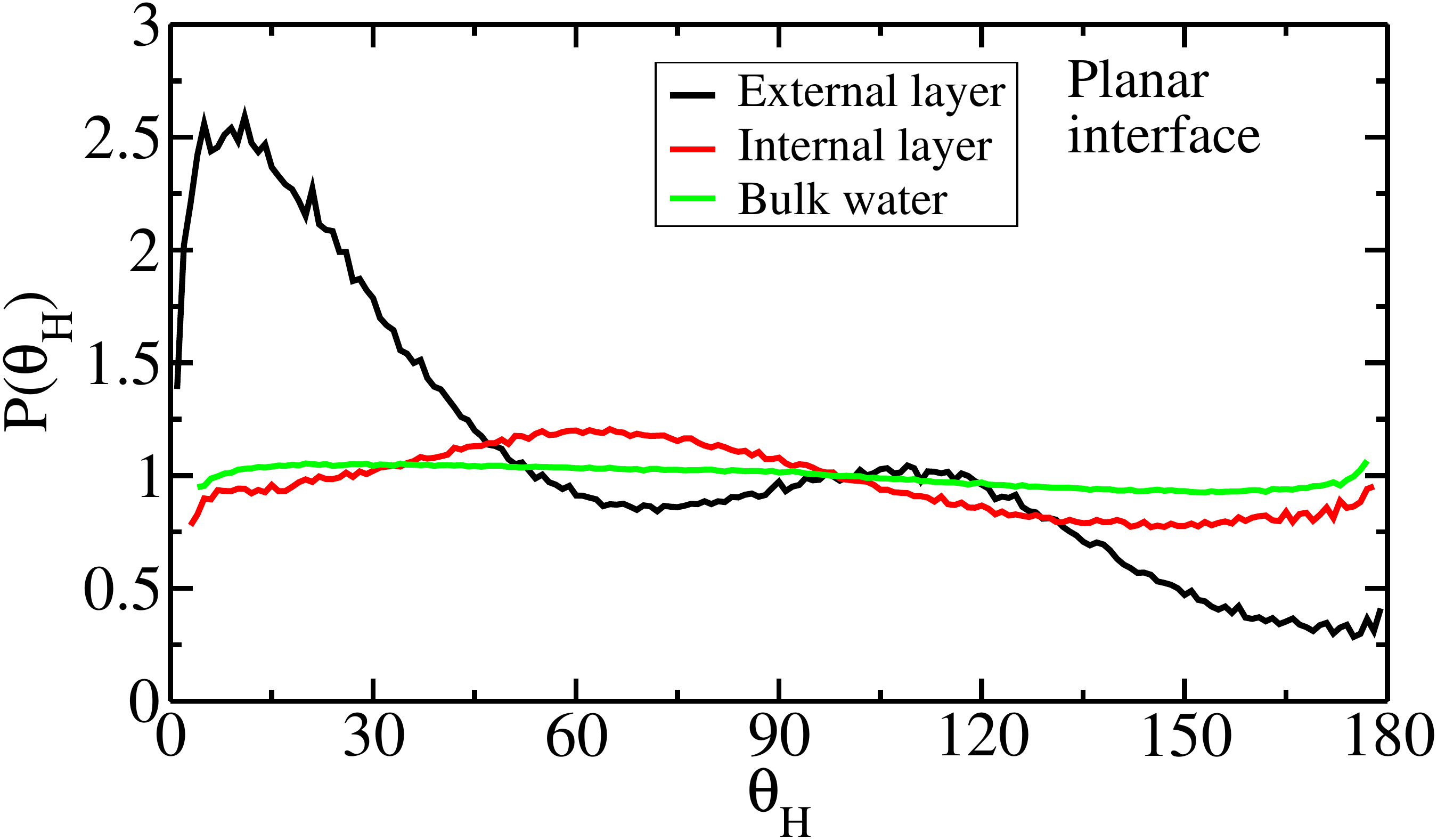} \includegraphics[width=0.45\linewidth]{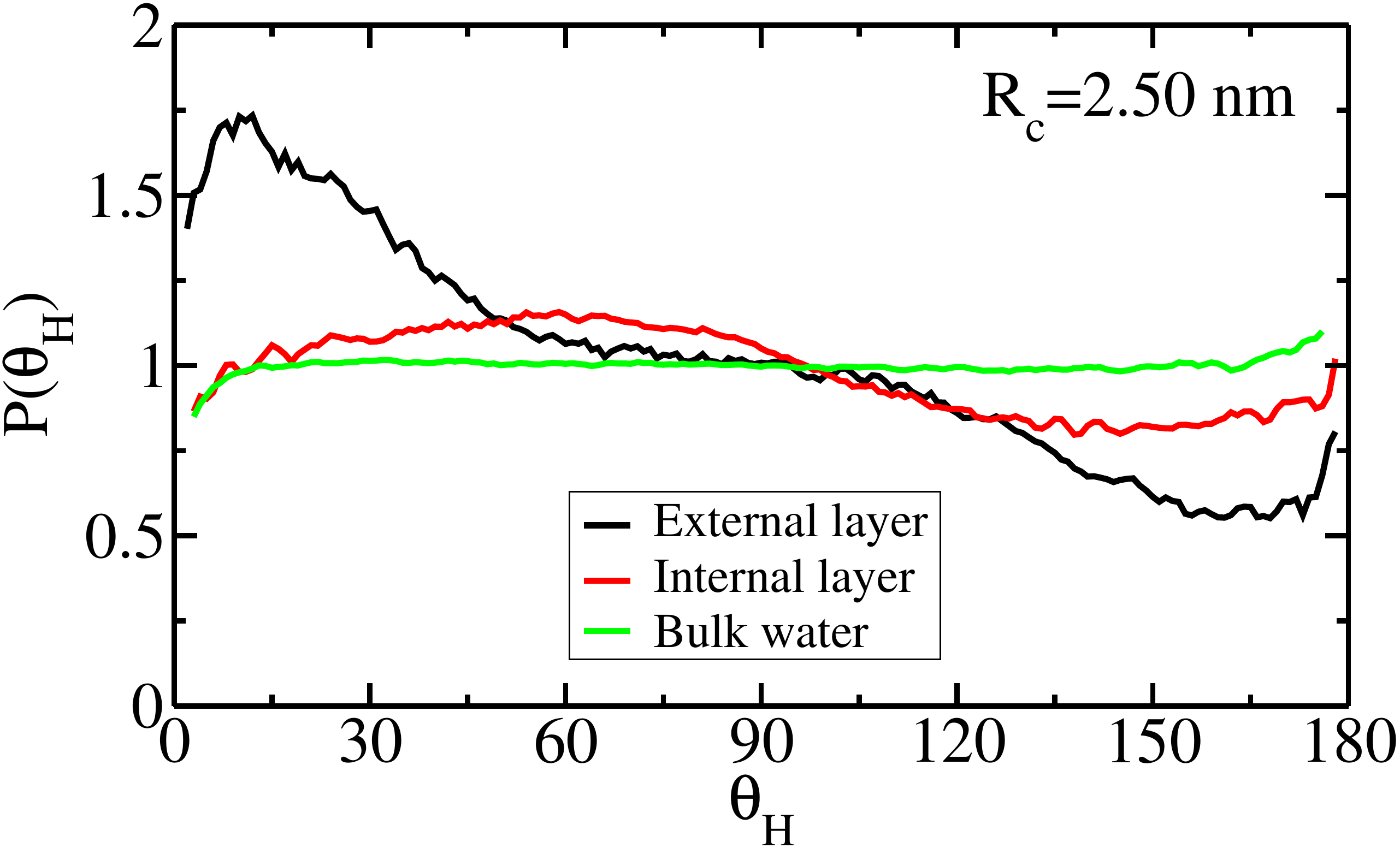} \\
    (c) \hspace{0.42\linewidth} (d) \\
    \includegraphics[width=0.45\linewidth]{normal_dpcur.pdf} \includegraphics[width=0.45\linewidth]{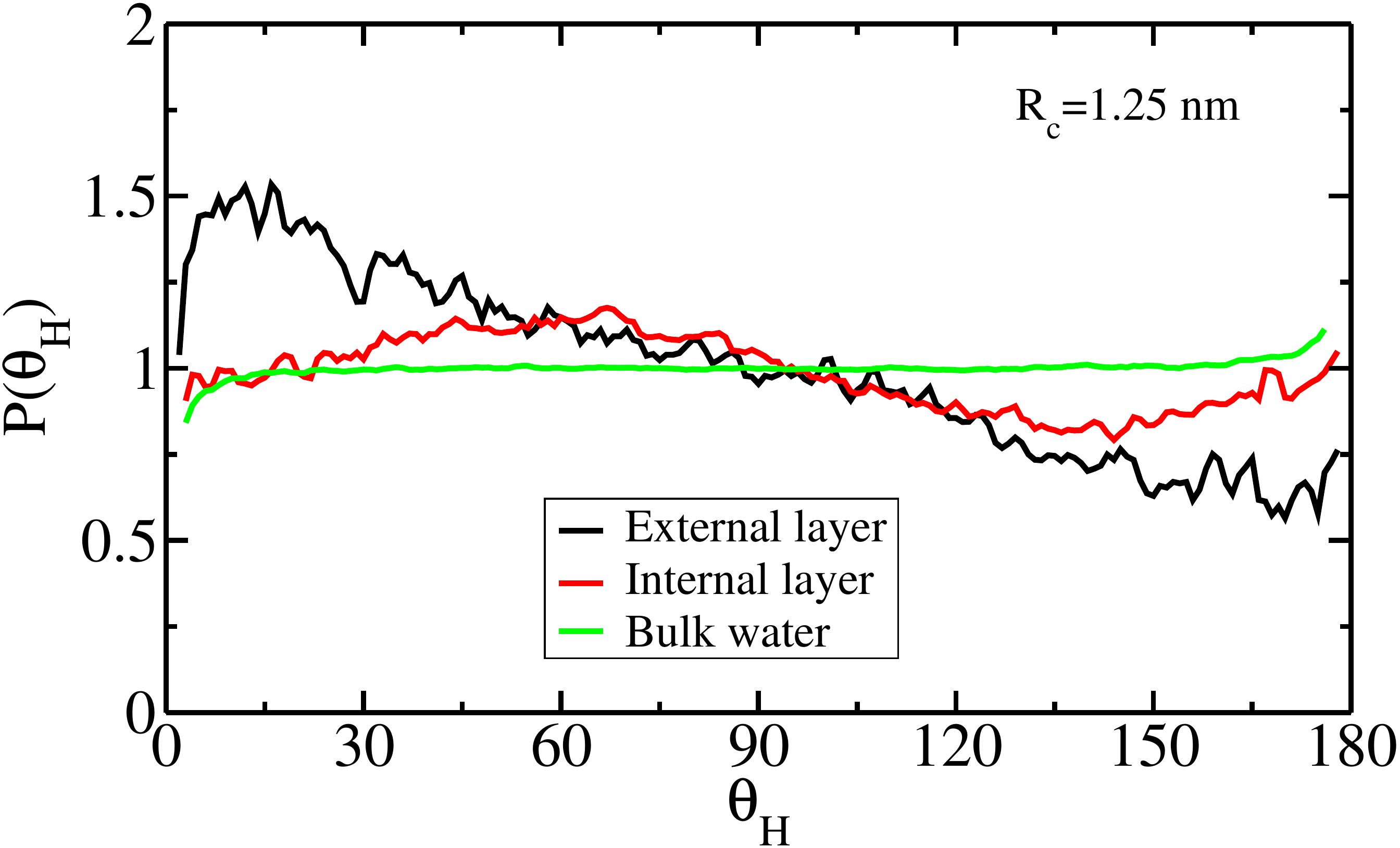}
    \caption{Distribution probability of the O-H bond with respect to the direction normal to the interface, renormalized by the random distribution sin($\theta _H$). The shown distributions are for (a) planar interface; (b-d) critical bubbles. The bubble radius is indicated in the leyend. }
    \label{curvature}
\end{figure}

In the main text we show the orientational analysis of interfacial molecules for one bubble size. In Figure \ref{curvature}, we show the same analysis for different bubble size using the DPMD model. We observe how the dependency of orientation preference with curvature is moderate but noticeable for the range of bubble sizes studied, being the biggest bubble the one resembling more to the coexistence conditions.

\bibliographystyle{ieeetr}

\begin{thebibliography}{100}

\bibitem{chanda2009organic}
A.~Chanda and V.~V. Fokin, ``Organic synthesis “on water”,'' {\em Chemical
  reviews}, vol.~109, no.~2, pp.~725--748, 2009.

\bibitem{chaplin2006we}
M.~Chaplin, ``Do we underestimate the importance of water in cell biology?,''
  {\em Nature Reviews Molecular Cell Biology}, vol.~7, no.~11, pp.~861--866,
  2006.

\bibitem{lester2006reaction}
E.~Lester, P.~Blood, J.~Denyer, D.~Giddings, B.~Azzopardi, and M.~Poliakoff,
  ``Reaction engineering: The supercritical water hydrothermal synthesis of
  nano-particles,'' {\em The Journal of Supercritical Fluids}, vol.~37, no.~2,
  pp.~209--214, 2006.

\bibitem{ponomarenko2014ultrasonic}
A.~Ponomarenko, O.~Vincent, A.~Pietriga, H.~Cochard, {\'E}.~Badel, and
  P.~Marmottant, ``Ultrasonic emissions reveal individual cavitation bubbles in
  water-stressed wood,'' {\em Journal of the Royal Society Interface}, vol.~11,
  no.~99, p.~20140480, 2014.

\bibitem{wheeler2008transpiration}
T.~D. Wheeler and A.~D. Stroock, ``The transpiration of water at negative
  pressures in a synthetic tree,'' {\em Nature}, vol.~455, no.~7210,
  pp.~208--212, 2008.

\bibitem{adhikari2015mechanism}
U.~Adhikari, A.~Goliaei, and M.~L. Berkowitz, ``Mechanism of membrane poration
  by shock wave induced nanobubble collapse: A molecular dynamics study,'' {\em
  The Journal of Physical Chemistry B}, vol.~119, no.~20, pp.~6225--6234, 2015.

\bibitem{yu2012effect}
D.~Yu, B.~Liu, and B.~Wang, ``The effect of ultrasonic waves on the nucleation
  of pure water and degassed water,'' {\em Ultrasonics sonochemistry}, vol.~19,
  no.~3, pp.~459--463, 2012.

\bibitem{kumar2010study}
P.~Kumar and R.~Saini, ``Study of cavitation in hydro turbines—a review,''
  {\em Renewable and Sustainable Energy Reviews}, vol.~14, no.~1, pp.~374--383,
  2010.

\bibitem{ma2011adsorption}
J.~Ma, A.~Michaelides, D.~Alfe, L.~Schimka, G.~Kresse, and E.~Wang,
  ``Adsorption and diffusion of water on graphene from first principles,'' {\em
  Physical Review B}, vol.~84, no.~3, p.~033402, 2011.

\bibitem{abascal2005general}
J.~L. Abascal and C.~Vega, ``A general purpose model for the condensed phases
  of water: Tip4p/2005,'' {\em The Journal of chemical physics}, vol.~123,
  no.~23, p.~234505, 2005.

\bibitem{abascal2005potential}
J.~Abascal, E.~Sanz, R.~Garc{\'\i}a~Fern{\'a}ndez, and C.~Vega, ``A potential
  model for the study of ices and amorphous water: Tip4p/ice,'' {\em The
  Journal of chemical physics}, vol.~122, no.~23, p.~234511, 2005.

\bibitem{mahoney2001quantum}
M.~W. Mahoney and W.~L. Jorgensen, ``Quantum, intramolecular flexibility, and
  polarizability effects on the reproduction of the density anomaly of liquid
  water by simple potential functions,'' {\em The Journal of Chemical Physics},
  vol.~115, no.~23, pp.~10758--10768, 2001.

\bibitem{jorgensen1983comparison}
W.~L. Jorgensen, J.~Chandrasekhar, J.~D. Madura, R.~W. Impey, and M.~L. Klein,
  ``Comparison of simple potential functions for simulating liquid water,''
  {\em The Journal of chemical physics}, vol.~79, no.~2, pp.~926--935, 1983.

\bibitem{horn2005characterization}
H.~W. Horn, W.~C. Swope, and J.~W. Pitera, ``Characterization of the tip4p-ew
  water model: Vapor pressure and boiling point,'' {\em The Journal of chemical
  physics}, vol.~123, no.~19, p.~194504, 2005.

\bibitem{berendsen1987missing}
H.~Berendsen, J.~Grigera, and T.~Straatsma, ``The missing term in effective
  pair potentials,'' {\em Journal of Physical Chemistry}, vol.~91, no.~24,
  pp.~6269--6271, 1987.

\bibitem{mahoney2000five}
M.~W. Mahoney and W.~L. Jorgensen, ``A five-site model for liquid water and the
  reproduction of the density anomaly by rigid, nonpolarizable potential
  functions,'' {\em The Journal of chemical physics}, vol.~112, no.~20,
  pp.~8910--8922, 2000.

\bibitem{molinero2009water}
V.~Molinero and E.~B. Moore, ``Water modeled as an intermediate element between
  carbon and silicon,'' {\em The Journal of Physical Chemistry B}, vol.~113,
  no.~13, pp.~4008--4016, 2009.

\bibitem{hasegawa2011polarizable}
T.~Hasegawa and Y.~Tanimura, ``A polarizable water model for intramolecular and
  intermolecular vibrational spectroscopies,'' {\em The Journal of Physical
  Chemistry B}, vol.~115, no.~18, pp.~5545--5553, 2011.

\bibitem{ahlstrom1989molecular}
P.~Ahlstr{\"o}m, A.~Wallqvist, S.~Engstr{\"o}m, and B.~J{\"o}nsson, ``A
  molecular dynamics study of polarizable water,'' {\em Molecular Physics},
  vol.~68, no.~3, pp.~563--581, 1989.

\bibitem{wang2013systematic}
L.-P. Wang, T.~Head-Gordon, J.~W. Ponder, P.~Ren, J.~D. Chodera, P.~K. Eastman,
  T.~J. Martinez, and V.~S. Pande, ``Systematic improvement of a classical
  molecular model of water,'' {\em The Journal of Physical Chemistry B},
  vol.~117, no.~34, pp.~9956--9972, 2013.

\bibitem{babin2013development}
V.~Babin, C.~Leforestier, and F.~Paesani, ``Development of a “first
  principles” water potential with flexible monomers: Dimer potential energy
  surface, vrt spectrum, and second virial coefficient,'' {\em Journal of
  chemical theory and computation}, vol.~9, no.~12, pp.~5395--5403, 2013.

\bibitem{santra2007accuracy}
B.~Santra, A.~Michaelides, and M.~Scheffler, ``On the accuracy of
  density-functional theory exchange-correlation functionals for h bonds in
  small water clusters: Benchmarks approaching the complete basis set limit,''
  {\em The Journal of chemical physics}, vol.~127, no.~18, p.~184104, 2007.

\bibitem{lambros2020good}
E.~Lambros and F.~Paesani, ``How good are polarizable and flexible models for
  water: Insights from a many-body perspective,'' {\em The Journal of Chemical
  Physics}, vol.~153, no.~6, p.~060901, 2020.

\bibitem{horn2004development}
H.~W. Horn, W.~C. Swope, J.~W. Pitera, J.~D. Madura, T.~J. Dick, G.~L. Hura,
  and T.~Head-Gordon, ``Development of an improved four-site water model for
  biomolecular simulations: Tip4p-ew,'' {\em The Journal of chemical physics},
  vol.~120, no.~20, pp.~9665--9678, 2004.

\bibitem{benavides2016consensus}
A.~Benavides, J.~Aragones, and C.~Vega, ``Consensus on the solubility of nacl
  in water from computer simulations using the chemical potential route,'' {\em
  The Journal of Chemical Physics}, vol.~144, no.~12, p.~124504, 2016.

\bibitem{espinosa2016calculation}
J.~Espinosa, J.~Young, H.~Jiang, D.~Gupta, C.~Vega, E.~Sanz, P.~G. Debenedetti,
  and A.~Z. Panagiotopoulos, ``On the calculation of solubilities via direct
  coexistence simulations: Investigation of nacl aqueous solutions and
  lennard-jones binary mixtures,'' {\em The Journal of chemical physics},
  vol.~145, no.~15, p.~154111, 2016.

\bibitem{jiang2018forward}
H.~Jiang, A.~Haji-Akbari, P.~G. Debenedetti, and A.~Z. Panagiotopoulos,
  ``Forward flux sampling calculation of homogeneous nucleation rates from
  aqueous nacl solutions,'' {\em The Journal of chemical physics}, vol.~148,
  no.~4, p.~044505, 2018.

\bibitem{sanchez2022direct}
I.~Sanchez-Burgos and J.~R. Espinosa, ``Direct calculation of the planar
  nacl-aqueous solution interfacial free energy at the solubility limit,'' {\em
  arXiv preprint arXiv:2208.08322}, 2022.

\bibitem{lamas2022freezing}
C.~P. Lamas, C.~Vega, and E.~G. Noya, ``Freezing point depression of salt
  aqueous solutions using the madrid-2019 model,'' {\em The Journal of chemical
  physics}, vol.~156, no.~13, p.~134503, 2022.

\bibitem{blazquez2022madrid}
S.~Blazquez, M.~Conde, J.~Abascal, and C.~Vega, ``The madrid-2019 force field
  for electrolytes in water using tip4p/2005 and scaled charges: Extension to
  the ions f-, br-, i-, rb+, and cs+,'' {\em The Journal of Chemical Physics},
  vol.~156, no.~4, p.~044505, 2022.

\bibitem{bergonzo2015improved}
C.~Bergonzo and T.~E. Cheatham~III, ``Improved force field parameters lead to a
  better description of rna structure,'' {\em Journal of chemical theory and
  computation}, vol.~11, no.~9, pp.~3969--3972, 2015.

\bibitem{smith2015force}
M.~D. Smith, J.~S. Rao, E.~Segelken, and L.~Cruz, ``Force-field induced bias in
  the structure of a$\beta$21--30: A comparison of opls, amber, charmm, and
  gromos force fields,'' {\em Journal of Chemical Information and Modeling},
  vol.~55, no.~12, pp.~2587--2595, 2015.

\bibitem{piaggi2020phase}
P.~M. Piaggi and R.~Car, ``Phase equilibrium of liquid water and hexagonal ice
  from enhanced sampling molecular dynamics simulations,'' {\em The Journal of
  chemical physics}, vol.~152, no.~20, p.~204116, 2020.

\bibitem{sanz2013homogeneous}
E.~Sanz, C.~Vega, J.~Espinosa, R.~Caballero-Bernal, J.~Abascal, and
  C.~Valeriani, ``Homogeneous ice nucleation at moderate supercooling from
  molecular simulation,'' {\em Journal of the American Chemical Society},
  vol.~135, no.~40, pp.~15008--15017, 2013.

\bibitem{espinosa2016seeding}
J.~R. Espinosa, C.~Vega, C.~Valeriani, and E.~Sanz, ``Seeding approach to
  crystal nucleation,'' {\em The Journal of chemical physics}, vol.~144, no.~3,
  p.~034501, 2016.

\bibitem{niu2019temperature}
H.~Niu, Y.~I. Yang, and M.~Parrinello, ``Temperature dependence of homogeneous
  nucleation in ice,'' {\em Physical review letters}, vol.~122, no.~24,
  p.~245501, 2019.

\bibitem{friesner2005ab}
R.~A. Friesner, ``Ab initio quantum chemistry: Methodology and applications,''
  {\em Proceedings of the National Academy of Sciences}, vol.~102, no.~19,
  pp.~6648--6653, 2005.

\bibitem{mlynsky2014comparison}
V.~Mlynsky, P.~Banas, J.~Sponer, M.~W. van~der Kamp, A.~J. Mulholland, and
  M.~Otyepka, ``Comparison of ab initio, dft, and semiempirical qm/mm
  approaches for description of catalytic mechanism of hairpin ribozyme,'' {\em
  Journal of Chemical Theory and Computation}, vol.~10, no.~4, pp.~1608--1622,
  2014.

\bibitem{gerber2014ab}
R.~Gerber, D.~Shemesh, M.~Varner, J.~Kalinowski, and B.~Hirshberg, ``Ab initio
  and semi-empirical molecular dynamics simulations of chemical reactions in
  isolated molecules and in clusters,'' {\em Physical Chemistry Chemical
  Physics}, vol.~16, no.~21, pp.~9760--9775, 2014.

\bibitem{zhang2018deep}
L.~Zhang, J.~Han, H.~Wang, R.~Car, and E.~Weinan, ``Deep potential molecular
  dynamics: a scalable model with the accuracy of quantum mechanics,'' {\em
  Physical review letters}, vol.~120, no.~14, p.~143001, 2018.

\bibitem{zhang2020dp}
Y.~Zhang, H.~Wang, W.~Chen, J.~Zeng, L.~Zhang, H.~Wang, and E.~Weinan,
  ``Dp-gen: A concurrent learning platform for the generation of reliable deep
  learning based potential energy models,'' {\em Computer Physics
  Communications}, vol.~253, p.~107206, 2020.

\bibitem{zhang2021phase}
L.~Zhang, H.~Wang, R.~Car, and E.~Weinan, ``Phase diagram of a deep potential
  water model,'' {\em Physical review letters}, vol.~126, no.~23, p.~236001,
  2021.

\bibitem{sun2015strongly}
J.~Sun, A.~Ruzsinszky, and J.~P. Perdew, ``Strongly constrained and
  appropriately normed semilocal density functional,'' {\em Physical review
  letters}, vol.~115, no.~3, p.~036402, 2015.

\bibitem{sun2016accurate}
J.~Sun, R.~C. Remsing, Y.~Zhang, Z.~Sun, A.~Ruzsinszky, H.~Peng, Z.~Yang,
  A.~Paul, U.~Waghmare, X.~Wu, {\em et~al.}, ``Accurate first-principles
  structures and energies of diversely bonded systems from an efficient density
  functional,'' {\em Nature chemistry}, vol.~8, no.~9, pp.~831--836, 2016.

\bibitem{piaggi2022pnas}
P.~M. Piaggi, J.~Weis, A.~Z. Panagiotopoulos, P.~G. Debenedetti, and R.~Car,
  ``Homogeneous ice nucleation in an ab initio machine-learning model of
  water,'' {\em Proceedings of the National Academy of Sciences}, vol.~119,
  no.~33, p.~e2207294119, 2022.

\bibitem{han2017deep}
J.~Han, L.~Zhang, R.~Car, {\em et~al.}, ``Deep potential: A general
  representation of a many-body potential energy surface,'' {\em arXiv preprint
  arXiv:1707.01478}, 2017.

\bibitem{Linstrom}
P.~Linstrom and W.~Mallard, {\em {NIST Chemistry WebBook, NIST Standard
  Reference Database Number 69}}.
\newblock Gaithersburg MD: National Institute of Standards and Technology.

\bibitem{vega2006vapor}
C.~Vega, J.~Abascal, and I.~Nezbeda, ``Vapor-liquid equilibria from the triple
  point up to the critical point for the new generation of tip4p-like models:
  Tip4p/ew, tip4p/2005, and tip4p/ice,'' {\em The Journal of chemical physics},
  vol.~125, no.~3, p.~034503, 2006.

\bibitem{vega2007surface}
C.~Vega and E.~de~Miguel, ``Surface tension of the most popular models of water
  by using the test-area simulation method,'' {\em The Journal of chemical
  physics}, vol.~126, no.~15, p.~154707, 2007.

\bibitem{mountain2009internally}
R.~D. Mountain, ``An internally consistent method for the molecular dynamics
  simulation of the surface tension: application to some tip4p-type models of
  water,'' {\em The Journal of Physical Chemistry B}, vol.~113, no.~2,
  pp.~482--486, 2009.

\bibitem{alejandre2010surface}
J.~Alejandre and G.~A. Chapela, ``The surface tension of tip4p/2005 water model
  using the ewald sums for the dispersion interactions,'' {\em The Journal of
  chemical physics}, vol.~132, no.~1, p.~014701, 2010.

\bibitem{pi2009anomalies}
H.~L. Pi, J.~L. Aragones, C.~Vega, E.~G. Noya, J.~L. Abascal, M.~A. Gonzalez,
  and C.~McBride, ``Anomalies in water as obtained from computer simulations of
  the tip4p/2005 model: density maxima, and density, isothermal compressibility
  and heat capacity minima,'' {\em Molecular Physics}, vol.~107, no.~4-6,
  pp.~365--374, 2009.

\bibitem{menzl2016molecular}
G.~Menzl, M.~A. Gonzalez, P.~Geiger, F.~Caupin, J.~L. Abascal, C.~Valeriani,
  and C.~Dellago, ``Molecular mechanism for cavitation in water under
  tension,'' {\em Proceedings of the National Academy of Sciences}, vol.~113,
  no.~48, pp.~13582--13587, 2016.

\bibitem{min2019bubbles}
S.~H. Min and M.~L. Berkowitz, ``Bubbles in water under stretch-induced
  cavitation,'' {\em The Journal of Chemical Physics}, vol.~150, no.~5,
  p.~054501, 2019.

\bibitem{abascal2013homogeneous}
J.~L. Abascal, M.~A. Gonzalez, J.~L. Aragones, and C.~Valeriani, ``Homogeneous
  bubble nucleation in water at negative pressure: A voronoi polyhedra
  analysis,'' {\em The Journal of chemical physics}, vol.~138, no.~8,
  p.~084508, 2013.

\bibitem{joswiak2013size}
M.~N. Joswiak, N.~Duff, M.~F. Doherty, and B.~Peters, ``Size-dependent surface
  free energy and tolman-corrected droplet nucleation of tip4p/2005 water,''
  {\em The Journal of Physical Chemistry Letters}, vol.~4, no.~24,
  pp.~4267--4272, 2013.

\bibitem{joswiak2016energetic}
M.~N. Joswiak, R.~Do, M.~F. Doherty, and B.~Peters, ``Energetic and entropic
  components of the tolman length for mw and tip4p/2005 water nanodroplets,''
  {\em The Journal of chemical physics}, vol.~145, no.~20, p.~204703, 2016.

\bibitem{gonzalez2015bubble}
M.~A. Gonzalez, J.~L. Abascal, C.~Valeriani, and F.~Bresme, ``Bubble nucleation
  in simple and molecular liquids via the largest spherical cavity method,''
  {\em The Journal of Chemical Physics}, vol.~142, no.~15, p.~154903, 2015.

\bibitem{gonzalez2014detecting}
M.~A. Gonz{\'a}lez, G.~Menzl, J.~L. Aragones, P.~Geiger, F.~Caupin, J.~L.
  Abascal, C.~Dellago, and C.~Valeriani, ``Detecting vapour bubbles in
  simulations of metastable water,'' {\em The Journal of chemical physics},
  vol.~141, no.~18, p.~18C511, 2014.

\bibitem{debenedetti2021metastable}
P.~G. Debenedetti, {\em Metastable liquids}.
\newblock Princeton university press, 2021.

\bibitem{kashchiev2000nucleation}
D.~Kashchiev, {\em Nucleation}.
\newblock Elsevier, 2000.

\bibitem{green1990water}
J.~Green, D.~Durben, G.~Wolf, and C.~Angell, ``Water and solutions at negative
  pressure: Raman spectroscopic study to-80 megapascals,'' {\em Science},
  vol.~249, no.~4969, pp.~649--652, 1990.

\bibitem{zheng1991liquids}
Q.~Zheng, D.~Durben, G.~Wolf, and C.~Angell, ``Liquids at large negative
  pressures: water at the homogeneous nucleation limit,'' {\em Science},
  vol.~254, no.~5033, pp.~829--832, 1991.

\bibitem{alvarenga1993elastic}
A.~Alvarenga, M.~Grimsditch, and R.~Bodnar, ``Elastic properties of water under
  negative pressures,'' {\em The Journal of chemical physics}, vol.~98, no.~11,
  pp.~8392--8396, 1993.

\bibitem{azouzi2013coherent}
M.~E.~M. Azouzi, C.~Ramboz, J.-F. Lenain, and F.~Caupin, ``A coherent picture
  of water at extreme negative pressure,'' {\em Nature Physics}, vol.~9, no.~1,
  pp.~38--41, 2013.

\bibitem{pallares2014anomalies}
G.~Pallares, M.~El~Mekki~Azouzi, M.~A. Gonz{\'a}lez, J.~L. Aragones, J.~L.
  Abascal, C.~Valeriani, and F.~Caupin, ``Anomalies in bulk supercooled water
  at negative pressure,'' {\em Proceedings of the National Academy of
  Sciences}, vol.~111, no.~22, pp.~7936--7941, 2014.

\bibitem{tolman1949effect}
R.~C. Tolman, ``The effect of droplet size on surface tension,'' {\em The
  journal of chemical physics}, vol.~17, no.~3, pp.~333--337, 1949.

\bibitem{salzmann2011polymorphism}
C.~G. Salzmann, P.~G. Radaelli, B.~Slater, and J.~L. Finney, ``The polymorphism
  of ice: five unresolved questions,'' {\em Physical Chemistry Chemical
  Physics}, vol.~13, no.~41, pp.~18468--18480, 2011.

\bibitem{wagner2011new}
W.~Wagner, T.~Riethmann, R.~Feistel, and A.~H. Harvey, ``New equations for the
  sublimation pressure and melting pressure of h2o ice ih,'' {\em Journal of
  Physical and Chemical Reference Data}, vol.~40, no.~4, p.~043103, 2011.

\bibitem{brown1966preliminary}
A.~Brown and E.~Whalley, ``Preliminary investigation of the phase boundaries
  between ice vi and vii and ice vi and viii,'' {\em The Journal of Chemical
  Physics}, vol.~45, no.~11, pp.~4360--4361, 1966.

\bibitem{lu2022dp}
D.~Lu, W.~Jiang, Y.~Chen, L.~Zhang, W.~Jia, H.~Wang, and M.~Chen, ``Dp
  compress: A model compression scheme for generating efficient deep potential
  models,'' {\em Journal of Chemical Theory and Computation}, vol.~18, no.~9,
  pp.~5559--5567, 2022.

\bibitem{Plimpton1995FastDynamics}
S.~Plimpton, ``{Fast parallel algorithms for short-range molecular dynamics},''
  {\em Journal of Computational Physics}, vol.~117, pp.~1--19, 3 1995.

\bibitem{wang2018deepmd}
H.~Wang, L.~Zhang, J.~Han, and E.~Weinan, ``Deepmd-kit: A deep learning package
  for many-body potential energy representation and molecular dynamics,'' {\em
  Computer Physics Communications}, vol.~228, pp.~178--184, 2018.

\bibitem{nosethermo}
S.~Nosé, ``A unified formulation of the constant temperature molecular
  dynamics methods,'' {\em The Journal of Chemical Physics}, vol.~81, no.~1,
  pp.~511--519, 1984.

\bibitem{hooverthermo}
W.~G. Hoover, ``Canonical dynamics: Equilibrium phase-space distributions,''
  {\em Phys. Rev. A}, vol.~31, pp.~1695--1697, Mar 1985.

\bibitem{hoover1986constant}
W.~G. Hoover, ``Constant-pressure equations of motion,'' {\em Physical Review
  A}, vol.~34, no.~3, p.~2499, 1986.

\bibitem{kirkwood1949statistical}
J.~G. Kirkwood and F.~P. Buff, ``The statistical mechanical theory of surface
  tension,'' {\em The Journal of Chemical Physics}, vol.~17, no.~3,
  pp.~338--343, 1949.

\bibitem{bekker1993gromacs}
H.~Bekker, H.~Berendsen, E.~Dijkstra, S.~Achterop, R.~Vondrumen,
  D.~VANDERSPOEL, A.~Sijbers, H.~Keegstra, and M.~Renardus, Gromacs-a
  parallel computer for molecular-dynamics simulations,'' in {\em 4th
  International Conference on Computational Physics (PC 92)}, pp.~252--256,
  World Scientific Publishing, 1993.

\bibitem{bussi2007canonical}
G.~Bussi, D.~Donadio, and M.~Parrinello, ``Canonical sampling through velocity
  rescaling,'' {\em The Journal of chemical physics}, vol.~126, no.~1,
  p.~014101, 2007.

\bibitem{parrinello1981polymorphic}
M.~Parrinello and A.~Rahman, ``Polymorphic transitions in single crystals: A
  new molecular dynamics method,'' {\em Journal of Applied physics}, vol.~52,
  no.~12, pp.~7182--7190, 1981.

\bibitem{hockney1974quiet}
R.~W. Hockney, S.~Goel, and J.~Eastwood, ``Quiet high-resolution computer
  models of a plasma,'' {\em Journal of Computational Physics}, vol.~14, no.~2,
  pp.~148--158, 1974.

\bibitem{darden1993particle}
T.~Darden, D.~York, and L.~Pedersen, Particle mesh ewald: An n log (n)
  method for ewald sums in large systems,'' {\em The Journal of chemical
  physics}, vol.~98, no.~12, pp.~10089--10092, 1993.

\bibitem{essmann1995smooth}
U.~Essmann, L.~Perera, M.~L. Berkowitz, T.~Darden, H.~Lee, and L.~G. Pedersen,
  ``A smooth particle mesh ewald method,'' {\em The Journal of chemical
  physics}, vol.~103, no.~19, pp.~8577--8593, 1995.

\bibitem{hess1997lincs}
B.~Hess, H.~Bekker, H.~J. Berendsen, and J.~G. Fraaije, ``Lincs: a linear
  constraint solver for molecular simulations,'' {\em Journal of computational
  chemistry}, vol.~18, no.~12, pp.~1463--1472, 1997.

\bibitem{rosales2020seeding}
P.~Rosales-Pelaez, I.~Sanchez-Burgos, C.~Valeriani, C.~Vega, and E.~Sanz,
  ``Seeding approach to nucleation in the n v t ensemble: The case of bubble
  cavitation in overstretched lennard jones fluids,'' {\em Physical Review E},
  vol.~101, no.~2, p.~022611, 2020.

\bibitem{montero2020interfacial}
P.~Montero~de Hijes, J.~R. Espinosa, V.~Bianco, E.~Sanz, and C.~Vega,
  ``Interfacial free energy and tolman length of curved liquid--solid
  interfaces from equilibrium studies,'' {\em The Journal of Physical Chemistry
  C}, vol.~124, no.~16, pp.~8795--8805, 2020.

\bibitem{volmer1926keimbildung}
M.~Volmer and A.~Weber, ``Keimbildung in {\"u}bers{\"a}ttigten gebilden,'' {\em
  Zeitschrift f{\"u}r physikalische Chemie}, vol.~119, no.~1, pp.~277--301,
  1926.

\bibitem{becker1935kinetische}
R.~Becker and W.~D{\"o}ring, ``Kinetische behandlung der keimbildung in
  {\"u}bers{\"a}ttigten d{\"a}mpfen,'' {\em Annalen der physik}, vol.~416,
  no.~8, pp.~719--752, 1935.

\bibitem{sear2012non}
R.~P. Sear, ``The non-classical nucleation of crystals: microscopic mechanisms
  and applications to molecular crystals, ice and calcium carbonate,'' {\em
  International Materials Reviews}, vol.~57, no.~6, pp.~328--356, 2012.

\bibitem{merikanto2007origin}
J.~Merikanto, E.~Zapadinsky, A.~Lauri, and H.~Vehkam{\"a}ki, ``Origin of the
  failure of classical nucleation theory: Incorrect description of the smallest
  clusters,'' {\em Physical review letters}, vol.~98, no.~14, p.~145702, 2007.

\bibitem{cacciuto2004breakdown}
A.~Cacciuto, S.~Auer, and D.~Frenkel, ``Breakdown of classical nucleation
  theory near isostructural phase transitions,'' {\em Physical review letters},
  vol.~93, no.~16, p.~166105, 2004.

\bibitem{horsch2008modification}
M.~Horsch, J.~Vrabec, and H.~Hasse, ``Modification of the classical nucleation
  theory based on molecular simulation data for surface tension, critical
  nucleus size, and nucleation rate,'' {\em Physical Review E}, vol.~78, no.~1,
  p.~011603, 2008.

\bibitem{moroni2005interplay}
D.~Moroni, P.~R. Ten~Wolde, and P.~G. Bolhuis, ``Interplay between structure
  and size in a critical crystal nucleus,'' {\em Physical review letters},
  vol.~94, no.~23, p.~235703, 2005.

\bibitem{leoni2021nonclassical}
F.~Leoni and J.~Russo, ``Nonclassical nucleation pathways in
  stacking-disordered crystals,'' {\em Physical Review X}, vol.~11, no.~3,
  p.~031006, 2021.

\bibitem{espinosa2018homogeneous}
J.~R. Espinosa, C.~Vega, and E.~Sanz, ``Homogeneous ice nucleation rate in
  water droplets,'' {\em The Journal of Physical Chemistry C}, vol.~122,
  no.~40, pp.~22892--22896, 2018.

\bibitem{knott2012homogeneous}
B.~C. Knott, V.~Molinero, M.~F. Doherty, and B.~Peters, ``Homogeneous
  nucleation of methane hydrates: Unrealistic under realistic conditions,''
  {\em Journal of the American Chemical Society}, vol.~134, no.~48,
  pp.~19544--19547, 2012.

\bibitem{lamas2021homogeneous}
C.~Lamas, J.~Espinosa, M.~Conde, J.~Ram{\'\i}rez, P.~M. de~Hijes, E.~G. Noya,
  C.~Vega, and E.~Sanz, ``Homogeneous nucleation of nacl in supersaturated
  solutions,'' {\em Physical Chemistry Chemical Physics}, vol.~23, no.~47,
  pp.~26843--26852, 2021.

\bibitem{saika2011nucleation}
I.~Saika-Voivod, F.~Romano, and F.~Sciortino, ``Nucleation barriers in
  tetrahedral liquids spanning glassy and crystallizing regimes,'' {\em The
  Journal of chemical physics}, vol.~135, no.~12, p.~124506, 2011.

\bibitem{richard2018crystallization}
D.~Richard and T.~Speck, ``Crystallization of hard spheres revisited. ii.
  thermodynamic modeling, nucleation work, and the surface of tension,'' {\em
  The Journal of chemical physics}, vol.~148, no.~22, p.~224102, 2018.

\bibitem{separdar2021molecular}
L.~Separdar, J.~P. Rino, and E.~D. Zanotto, ``Molecular dynamics simulations of
  spontaneous and seeded nucleation and theoretical calculations for zinc
  selenide,'' {\em Computational Materials Science}, vol.~187, p.~110124, 2021.

\bibitem{sanchez2021parasitic}
I.~Sanchez-Burgos, A.~Garaizar, C.~Vega, E.~Sanz, and J.~R. Espinosa,
  ``Parasitic crystallization of colloidal electrolytes: growing a metastable
  crystal from the nucleus of a stable phase,'' {\em Soft Matter}, vol.~17,
  no.~3, pp.~489--505, 2021.

\bibitem{sanchez2021fcc}
I.~Sanchez-Burgos, E.~Sanz, C.~Vega, and J.~R. Espinosa, ``Fcc vs. hcp
  competition in colloidal hard-sphere nucleation: on their relative stability,
  interfacial free energy and nucleation rate,'' {\em Physical Chemistry
  Chemical Physics}, vol.~23, no.~35, pp.~19611--19626, 2021.

\bibitem{filion2010crystal}
L.~Filion, M.~Hermes, R.~Ni, and M.~Dijkstra, ``Crystal nucleation of hard
  spheres using molecular dynamics, umbrella sampling, and forward flux
  sampling: A comparison of simulation techniques,'' {\em The Journal of
  chemical physics}, vol.~133, no.~24, p.~244115, 2010.

\bibitem{fiorucci2020effect}
G.~Fiorucci, G.~M. Coli, J.~T. Padding, and M.~Dijkstra, ``The effect of
  hydrodynamics on the crystal nucleation of nearly hard spheres,'' {\em The
  Journal of Chemical Physics}, vol.~152, no.~6, p.~064903, 2020.

\bibitem{piaggipnas}
P.~M. Piaggi, J.~Weis, A.~Z. Panagiotopoulos, P.~G. Debenedetti, and R.~Car,
  ``Homogeneous ice nucleation in an ab initio machine-learning model of
  water,'' {\em Proceedings of the National Academy of Sciences}, vol.~119,
  no.~33, p.~e2207294119, 2022.

\bibitem{baidakov2022stability}
V.~Baidakov, ``Stability of metastable phases and kinetics of nucleation in a
  simple single-component system (molecular dynamics simulation)(a review),''
  {\em Russian Journal of General Chemistry}, vol.~92, no.~4, pp.~611--628,
  2022.

\bibitem{sanchez2020equivalence}
I.~Sanchez-Burgos, P.~M. de~Hijes, P.~Rosales-Pelaez, C.~Vega, and E.~Sanz,
  ``Equivalence between condensation and boiling in a lennard-jones fluid,''
  {\em Physical Review E}, vol.~102, no.~6, p.~062609, 2020.

\bibitem{blander1975bubble}
M.~Blander and J.~L. Katz, ``Bubble nucleation in liquids,'' {\em AIChE
  Journal}, vol.~21, no.~5, pp.~833--848, 1975.

\bibitem{bal2022extending}
K.~M. Bal and E.~C. Neyts, ``Extending and validating bubble nucleation rate
  predictions in a lennard-jones fluid with enhanced sampling methods and
  transition state theory,'' {\em The Journal of Chemical Physics}, vol.~157,
  no.~18, p.~184113, 2022.

\bibitem{diemand2014direct}
J.~Diemand, R.~Ang{\'e}lil, K.~K. Tanaka, and H.~Tanaka, ``Direct simulations
  of homogeneous bubble nucleation: Agreement with classical nucleation theory
  and no local hot spots,'' {\em Physical review E}, vol.~90, no.~5, p.~052407,
  2014.

\bibitem{montero2020young}
P.~Montero~de Hijes, K.~Shi, E.~G. Noya, E.~Santiso, K.~Gubbins, E.~Sanz, and
  C.~Vega, ``The young--laplace equation for a solid--liquid interface,'' {\em
  The Journal of Chemical Physics}, vol.~153, no.~19, p.~191102, 2020.

\bibitem{meadley2012thermodynamics}
S.~L. Meadley and F.~A. Escobedo, ``Thermodynamics and kinetics of bubble
  nucleation: Simulation methodology,'' {\em The Journal of chemical physics},
  vol.~137, no.~7, p.~074109, 2012.

\bibitem{kusaka1999identifying}
I.~Kusaka and D.~W. Oxtoby, ``Identifying physical clusters in bubble
  nucleation,'' {\em The Journal of chemical physics}, vol.~111, no.~3,
  pp.~1104--1108, 1999.

\bibitem{angelil2014bubble}
R.~Ang{\'e}lil, J.~Diemand, K.~K. Tanaka, and H.~Tanaka, ``Bubble evolution and
  properties in homogeneous nucleation simulations,'' {\em Physical review E},
  vol.~90, no.~6, p.~063301, 2014.

\bibitem{mohanchen2017aimdwater}
M.~Chen, H.-Y. Ko, R.~C. Remsing, M.~F.~C. Andrade, B.~Santra, Z.~Sun,
  A.~Selloni, R.~Car, M.~L. Klein, J.~P. Perdew, and X.~Wu, ``Ab initio theory
  and modeling of water,'' {\em Proceedings of the National Academy of
  Sciences}, vol.~114, no.~41, pp.~10846--10851, 2017.

\bibitem{yaoyi2020scanshift}
Y.~Yao and Y.~Kanai, ``Temperature dependence of nuclear quantum effects on
  liquid water via artificial neural network model based on scan meta-gga
  functional,'' {\em The Journal of Chemical Physics}, vol.~153, no.~4,
  p.~044114, 2020.

\bibitem{tom2020signatureLLT}
T.~E. Gartner, L.~Zhang, P.~M. Piaggi, R.~Car, A.~Z. Panagiotopoulos, and P.~G.
  Debenedetti, ``Signatures of a liquid–liquid transition in an ab initio
  deep neural network model for water,'' {\em Proceedings of the National
  Academy of Sciences}, vol.~117, no.~42, pp.~26040--26046, 2020.

\bibitem{piaggi2021phaseequilibrium}
P.~M. Piaggi, A.~Z. Panagiotopoulos, P.~G. Debenedetti, and R.~Car, ``Phase
  equilibrium of water with hexagonal and cubic ice using the scan
  functional,'' {\em Journal of Chemical Theory and Computation}, vol.~17,
  no.~5, pp.~3065--3077, 2021.
\newblock PMID: 33835819.

\bibitem{stukowski2009visualization}
A.~Stukowski, ``Visualization and analysis of atomistic simulation data with
  ovito--the open visualization tool,'' {\em Modelling and simulation in
  materials science and engineering}, vol.~18, no.~1, p.~015012, 2009.

\bibitem{rowlinson2013molecular}
J.~S. Rowlinson and B.~Widom, {\em Molecular theory of capillarity}.
\newblock Courier Corporation, USA, 2013.

\bibitem{zollweg1972law}
J.~A. Zollweg and G.~W. Mulholland, ``On the law of the rectilinear diameter,''
  {\em The Journal of Chemical Physics}, vol.~57, no.~3, pp.~1021--1025, 1972.

\bibitem{torrie1974monte}
G.~M. Torrie and J.~P. Valleau, ``Monte carlo free energy estimates using
  non-boltzmann sampling: Application to the sub-critical lennard-jones
  fluid,'' {\em Chemical Physics Letters}, vol.~28, no.~4, pp.~578--581, 1974.

\bibitem{thompson2009general}
A.~P. Thompson, S.~J. Plimpton, and W.~Mattson, ``General formulation of
  pressure and stress tensor for arbitrary many-body interaction potentials
  under periodic boundary conditions,'' {\em The Journal of chemical physics},
  vol.~131, no.~15, p.~154107, 2009.

\bibitem{van2009direct}
A.~E. van Giessen and E.~M. Blokhuis, ``Direct determination of the tolman
  length from the bulk pressures of liquid drops via molecular dynamics
  simulations,'' {\em The Journal of chemical physics}, vol.~131, no.~16,
  p.~164705, 2009.

\bibitem{bykov1999patching}
T.~Bykov and X.~C. Zeng, ``A patching model for surface tension and the tolman
  length,'' {\em The Journal of chemical physics}, vol.~111, no.~8,
  pp.~3705--3713, 1999.

\bibitem{lei2005tolman}
Y.~A. Lei, T.~Bykov, S.~Yoo, and X.~C. Zeng, ``The tolman length: Is it
  positive or negative?,'' {\em Journal of the American Chemical Society},
  vol.~127, no.~44, pp.~15346--15347, 2005.

\bibitem{julin2010thermodynamically}
J.~Julin, I.~Napari, J.~Merikanto, and H.~Vehkam{\"a}ki, ``A thermodynamically
  consistent determination of surface tension of small lennard-jones clusters
  from simulation and theory,'' {\em The Journal of chemical physics},
  vol.~133, no.~4, p.~044704, 2010.

\bibitem{kashchiev2020nucleation}
D.~Kashchiev, ``Nucleation work, surface tension, and gibbs--tolman length for
  nucleus of any size,'' {\em The Journal of Chemical Physics}, vol.~153,
  no.~12, p.~124509, 2020.

\bibitem{troster2011positive}
A.~Tr{\"o}ster and K.~Binder, ``Positive tolman length in a lattice gas with
  three-body interactions,'' {\em Physical Review Letters}, vol.~107, no.~26,
  p.~265701, 2011.

\bibitem{iwamatsu1997temperature}
M.~Iwamatsu and K.~Horii, ``Temperature dependence of liquid-vapor nucleation
  rate for the yukawa model fluid,'' {\em Aerosol science and technology},
  vol.~27, no.~5, pp.~563--574, 1997.

\bibitem{magaletti2021water}
F.~Magaletti, M.~Gallo, and C.~M. Casciola, ``Water cavitation from ambient to
  high temperatures,'' {\em Scientific reports}, vol.~11, no.~1, pp.~1--10,
  2021.

\bibitem{tang2020molecular}
F.~Tang, T.~Ohto, S.~Sun, J.~R. Rouxel, S.~Imoto, E.~H. Backus, S.~Mukamel,
  M.~Bonn, and Y.~Nagata, ``Molecular structure and modeling of water--air and
  ice--air interfaces monitored by sum-frequency generation,'' {\em Chemical
  reviews}, vol.~120, no.~8, pp.~3633--3667, 2020.

\bibitem{nagata2015surface}
Y.~Nagata, T.~Hasegawa, E.~H. Backus, K.~Usui, S.~Yoshimune, T.~Ohto, and
  M.~Bonn, ``The surface roughness, but not the water molecular orientation
  varies with temperature at the water--air interface,'' {\em Physical
  Chemistry Chemical Physics}, vol.~17, no.~36, pp.~23559--23564, 2015.

\bibitem{bonn2015molecular}
M.~Bonn, Y.~Nagata, and E.~H. Backus, ``Molecular structure and dynamics of
  water at the water--air interface studied with surface-specific vibrational
  spectroscopy,'' {\em Angewandte Chemie International Edition}, vol.~54,
  no.~19, pp.~5560--5576, 2015.

\bibitem{nihonyanagi2013structure}
S.~Nihonyanagi, J.~A. Mondal, S.~Yamaguchi, and T.~Tahara, ``Structure and
  dynamics of interfacial water studied by heterodyne-detected vibrational
  sum-frequency generation,'' {\em Annual review of physical chemistry},
  vol.~64, pp.~579--603, 2013.

\bibitem{nihonyanagi2011unified}
S.~Nihonyanagi, T.~Ishiyama, T.-k. Lee, S.~Yamaguchi, M.~Bonn, A.~Morita, and
  T.~Tahara, ``Unified molecular view of the air/water interface based on
  experimental and theoretical $\chi$ (2) spectra of an isotopically diluted
  water surface,'' {\em Journal of the American Chemical Society}, vol.~133,
  no.~42, pp.~16875--16880, 2011.

\bibitem{liang2019ab}
C.~Liang, J.~Jeon, and M.~Cho, ``Ab initio modeling of the vibrational
  sum-frequency generation spectrum of interfacial water,'' {\em The Journal of
  Physical Chemistry Letters}, vol.~10, no.~5, pp.~1153--1158, 2019.

\bibitem{ohto2015toward}
T.~Ohto, K.~Usui, T.~Hasegawa, M.~Bonn, and Y.~Nagata, ``Toward ab initio
  molecular dynamics modeling for sum-frequency generation spectra; an
  efficient algorithm based on surface-specific velocity-velocity correlation
  function,'' {\em The Journal of Chemical Physics}, vol.~143, no.~12,
  p.~124702, 2015.

\bibitem{kessler2015structure}
J.~Kessler, H.~Elgabarty, T.~Spura, K.~Karhan, P.~Partovi-Azar, A.~A.
  Hassanali, and T.~D. Kuhne, Structure and dynamics of the instantaneous
  water/vapor interface revisited by path-integral and ab initio molecular
  dynamics simulations,'' {\em The Journal of Physical Chemistry B}, vol.~119,
  no.~31, pp.~10079--10086, 2015.

\bibitem{vassilev2001ab}
P.~Vassilev, C.~Hartnig, M.~T. Koper, F.~Frechard, and R.~A. van Santen, ``Ab
  initio molecular dynamics simulation of liquid water and water--vapor
  interface,'' {\em The Journal of Chemical Physics}, vol.~115, no.~21,
  pp.~9815--9820, 2001.

\bibitem{khatib2016molecular}
R.~Khatib, T.~Hasegawa, M.~Sulpizi, E.~H. Backus, M.~Bonn, and Y.~Nagata,
  ``Molecular dynamics simulations of sfg librational modes spectra of water at
  the water--air interface,'' {\em The Journal of Physical Chemistry C},
  vol.~120, no.~33, pp.~18665--18673, 2016.

\bibitem{lee1984structure}
C.-Y. Lee, J.~A. McCammon, and P.~Rossky, ``The structure of liquid water at an
  extended hydrophobic surface,'' {\em The Journal of chemical physics},
  vol.~80, no.~9, pp.~4448--4455, 1984.

\bibitem{fan2009structure}
Y.~Fan, X.~Chen, L.~Yang, P.~S. Cremer, and Y.~Q. Gao, ``On the structure of
  water at the aqueous/air interface,'' {\em The Journal of Physical Chemistry
  B}, vol.~113, no.~34, pp.~11672--11679, 2009.

\bibitem{espinosa2016ice}
J.~R. Espinosa, C.~Vega, and E.~Sanz, ``Ice--water interfacial free energy for
  the tip4p, tip4p/2005, tip4p/ice, and mw models as obtained from the mold
  integration technique,'' {\em The Journal of Physical Chemistry C}, vol.~120,
  no.~15, pp.~8068--8075, 2016.

\end{thebibliography}

\begin{thebibliography}{1}

\bibitem{edelsbrunner1994threeKK}
H.~Edelsbrunner and E.~P. M{\"u}cke, ``Three-dimensional alpha shapes,'' {\em
  ACM Transactions on Graphics (TOG)}, vol.~13, no.~1, pp.~43--72, 1994.

\bibitem{stukowski2014computationalKK}
A.~Stukowski, ``Computational analysis methods in atomistic modeling of
  crystals,'' {\em Jom}, vol.~66, no.~3, pp.~399--407, 2014.

\bibitem{stukowski2009visualizationKK}
A.~Stukowski, ``Visualization and analysis of atomistic simulation data with
  ovito--the open visualization tool,'' {\em Modelling and Simulation in
  Materials Science and Engineering}, vol.~18, no.~1, p.~015012, 2009.

\bibitem{wadell1935volumeKK}
H.~Wadell, ``Volume, shape, and roundness of quartz particles,'' {\em The
  Journal of Geology}, vol.~43, no.~3, pp.~250--280, 1935.

\end{thebibliography}

\end{document}